# Uniformly curated signaling pathways reveal tissue-specific cross-talks and support drug target discovery

Tamás Korcsmáros[1,2,‡], Illés J. Farkas[3,‡], Máté S. Szalay[2], Petra Rovó[1], Dávid Fazekas[1], Zoltán Spiró[2], Csaba Böde[4], Katalin Lenti[5], Tibor Vellai[1], and Péter Csermely[2,*]

[1] Department of Genetics, Eötvös University, Pázmány P. s. 1C, H-1117 Budapest, Hungary

[2] Department of Medical Chemistry, Semmelweis University, PO Box 260, H-1444, Budapest, Hungary

[3] Statistical and Biological Physics Group of the Hungarian Acad. of Sciences, Pázmány P. s. 1A, H-1117 Budapest, Hungary

[4] Morgan Stanley Hungary Analytics Ltd., Lechner Ö. f. 8, H-1095 Budapest, Hungary

[5] Department of Morphology and Physiology, Semmelweis University, Vas u. 17, H-1088 Budapest, Hungary

[‡] equal contributions

**ABSTRACT**

**Motivation:** Signaling pathways control a large variety of cellular processes. However, currently, even within the same database signaling pathways are often curated at different levels of detail. This makes comparative and cross-talk analyses difficult.

**Results:** We present SignaLink, a database containing 8 major signaling pathways from *Caenorhabditis elegans*, *Drosophila melanogaster*, and humans. Based on 170 review and ~800 research articles, we have compiled pathways with semi-automatic searches and uniform, well-documented curation rules. We found that in humans any two of the 8 pathways can cross-talk. We quantified the possible tissue- and cancer-specific activity of cross-talks and found pathway-specific expression profiles. In addition, we identified 327 proteins relevant for drug target discovery.

**Conclusions:** We provide a novel resource for comparative and cross-talk analyses of signaling pathways. The identified multi-pathway and tissue-specific cross-talks contribute to the understanding of the signaling complexity in health and disease and underscore its importance in network-based drug target selection.

**Availability:** http://SignaLink.org

## 1 INTRODUCTION

Intracellular signaling, from the simplest cascades to the highly intertwined networks of protein kinases, contributes extensively to the diversity of developmental programs and adaptation responses in metazoans (Pires-daSilva and Sommer, 2003). In humans, defects in intracellular signaling can cause various diseases, *e.g.*, cancer, neurodegeneration, or diabetes. Thus, understanding the structure, function, and evolution of signal transduction is an important task for both basic research and medicine. By now genetic studies have uncovered functionally separate, though interacting (cross-talking), pathways and the direction of information flow between pairs of signaling molecules in a number of species (Beyer et al., 2007). On the other hand, biochemical experiments have allowed the detailed characterization of direct physical interactions involved in signaling (Xia et al., 2004). Integrating these data sets using uniform manual curation criteria can significantly contribute to a more precise assessment of their tissue- and cancer-specific utilization and the effects of drug treatments (Davidov et al., 2003). For example, inhibitors used for eliminating a signaling pathway in cancerous cells may in fact have the opposite effect. These drugs may suppress negative feedback loops and thereby, paradoxically, activate the targeted pathway (Sergina et al., 2007).

Intracellular signaling was originally regarded as an assembly of distinct and almost linear cascades. Over the past decade, however, it has been realized that signaling pathways are highly structured and rich in cross-talks (where cross-talk is defined here as a directed physical interaction between pathways). Consequently, intracellular signaling is now viewed as a set of intertwined pathways forming a single signaling network (Papin et al., 2005). This paradigm shift calls for novel experimental, curation, and network modeling techniques (Bauer-Mehren et al., 2009).

Currently, high-throughput (HTP) experiments are the major sources of known protein-protein interactions. However, so far in most HTP experiments extracellular, membrane-bound, and nuclear proteins have been underrepresented. These and other sampling biases strongly reduce their usability for identifying signaling interactions. Another limitation of HTP assays is that they produce undirected interactions even though in signaling directions are essential. Accordingly, several signaling pathway databases have been created recently by manually collecting the directed interactions from the literature (Bauer-Mehren et al., 2009).

Manually curated signaling pathway databases are often assembled without strictly defined and published standardized curation criteria (Lu et al., 2007). Therefore, even within the same database, *e.g.,* in KEGG (Ogata et al., 1999), the level of detail of curation and the rules for setting pathway boundaries can vary among pathways. In addition, in several signaling resources the definition of signaling pathways has no evolutionary or biochemical background. In other cases, *e.g.*, in Reactome and NetPath (Kandasamy et al., 2010; Joshi-Tope et al., 2005), curation criteria are standardized, however, (i) pathways are usually handled as separate entities, (ii) cross-talks and multi-pathway proteins are underrepresented, and (iii) extracting signaling information from the databases are complicated and labor-intensive, see Discussion and Supplementary Data for details. Another limitation of several current signaling resources is that they neglect the importance of multi-pathway proteins, *i.e.*, proteins functioning in more than one pathway (Komarova et al., 2005). In summary, the manual curation process needs to be uniform across all pathways and species to aid cross-talk analyses, tests of evolutionary hypotheses, dynamical modeling, setting up predictions, and drug target selection (Table 1).

* To whom correspondence should be addressed.
E-mail: csermely@eok.sote.hu

We present SignaLink, a signaling resource compiled by applying uniform manual curation rules and data structures across 8 major, biochemically defined signaling pathways in 3 metazoans (Fig. 1). The curation method allowed a systematic comparison of pathway sizes and cross-talks. We found that in humans any two of the 8 pathways can cross-talk and in humans we compared the possible dynamic activities of both the pathways and their cross-talks. We characterized tissue- and cancer-specific expression profiles, and identified proteins relevant for drug target discovery.

**Table 1.** Comparison of the curation processes of 3 manually curated databases and SignaLink. Only the differences from SignaLink are listed. The features setting SignaLink most clearly apart have a grey background.

| | KEGG | Reactome | NetPath | SignaLink |
|---|---|---|---|---|
| **Source of Signaling Proteins** | Selective manual curation from the literature | Reaction-based manual curation | Selective manual curation from HPRD | Manual curation: pathway-based reviews, experimental papers |
| **Source of Interactions** | Only reviews | PubMed ID available for each interaction | PubMed ID reaction details for each interaction | PubMed ID listed for each interaction |
| **Number of Signaling Pathways** | 10 (no NHR) | 10, but not for all species (no Hh ,NHR, JAK/STAT) | 20 (no WNT, IGF, NHR, JAK/STAT) | 8 (EGF/MAPK, WNT, TGF, Notch, IGF, Hh, NHR, JAK/STAT) |
| **Definitions of Pathways** | Not available | Not available; uniform curation rules | Cancer or immune pathways | Biochemically defined; important in development; uniform curation rules; documented in detail |
| **Pathways in one platform for cross-talk analysis** | Not possible | Possible, but no global pathway view or common platform | Not possible | Possible |

## 2 SYSTEM AND METHODS

### 2.1 Signaling proteins and interactions

SignaLink lists signaling proteins and directed signaling interactions between pairs of proteins in healthy cells of *Caenorhabditis elegans*, *Drosophila melanogaster*, and *Homo sapiens*. Each interaction is documented with the PubMed ID of the publication reporting the verifying experiment(s). SignaLink was compiled separately for all pathways of the 3 organisms. Search functions, data, and network images of the pathways are available at http://SignaLink.org.

In each of the 3 organisms we first listed signaling proteins and interactions from reviews (and from WormBook in *C. elegans*) and then added further signaling interactions of the listed proteins. To identify additional interactions in *C. elegans* we examined all interactions (except for transcription regulation) of the signaling proteins listed in WormBase (Rogers et al., 2008) and added only those to SignaLink that we could manually identify in the literature as an experimentally verified signaling interaction. For *D. melanogaster* we added to SignaLink those genetic interactions from FlyBase (Drysdale, 2008) that were also reported in at least one yeast-2-hybrid experiment. For humans we manually checked the reliability and directions for the protein-protein interactions found with the search engines iHop and Chilibot (Chen and Sharp, 2004; Hoffmann and Valencia, 2004).

SignaLink assigns proteins to signaling pathways using the full texts of pathway-reviews (written by pathway experts). While most signaling resources consider 5-15 reviews per pathway, SignaLink uses a total of 170 review papers, *i.e.*, more than 20 per pathway on average. Interactions were curated from a total of 941 articles (PubMed IDs are available at the website). We added a small number of proteins based on InParanoid ortholog clusters (Berglund et al., 2008). For curation we used a self-developed graphical tool and Perl/Python scripts. The current version of SignaLink was completed in May 2008 based on WormBase (version 191), FlyBase (2008.6), Ensembl (49), UniProt (87), and the publications listed on the website. Pathway data can be downloaded in several formats: SQL, CSV, XLS, CYS and SVG exported from Cytoscape, and SBML.

### 2.2 Quality control, database validation, statistical significance tests

The curation protocol of SignaLink (Fig. 1A) contains several steps aimed specifically at reducing data and curation errors. We used reviews as a starting point, manually looked up interactions three times, and manually searched for interactions of known signaling proteins with no signaling interactions so far in the database. The section “Advantages and limitations of SignaLink” explains validation steps and results in detail. We performed functional significance tests for each of the signaling pathways and their overlaps, *i.e.* multi-pathway proteins, with the ‘GO Termfinder’ toolbox (Boyle et al., 2004). We found a significant functional similarity between the functions of multi-pathway proteins and the functions of their pathways compared to the control case (functional similarity between the functions of all proteins and all pathways). Moreover, we statistically evaluated the human interactions listed in SignaLink with the PRINCESS web service (Li et al., 2008) and found that the ratio of high confidence interactions is 90.6%. Details for all statistical significance tests are available in the Supplementary Data.

### 2.3 Expression in selected healthy tissues and liver carcinomas

To investigate the dynamic activity of pathway interactions, we selected five healthy tissue types – colorectal, muscle, skin, liver, and cardiovascular tissues – and 2 liver carcinomas (Fig. 3, see the details for the selection process of the tissues and controls in the Supplementary Data file. Protein expression data in healthy tissue types were downloaded from the eGenetics database (integrated into Ensembl). Protein expression data in 2 screens of liver carcinomas were obtained from Oncomine 3.6 (Rhodes et al., 2007). We considered a protein differentially expressed if the p value of its expression in at least one of the 2 screens, as compared to healthy liver tissues and computed by a t-test of Oncomine, was below 0.05.

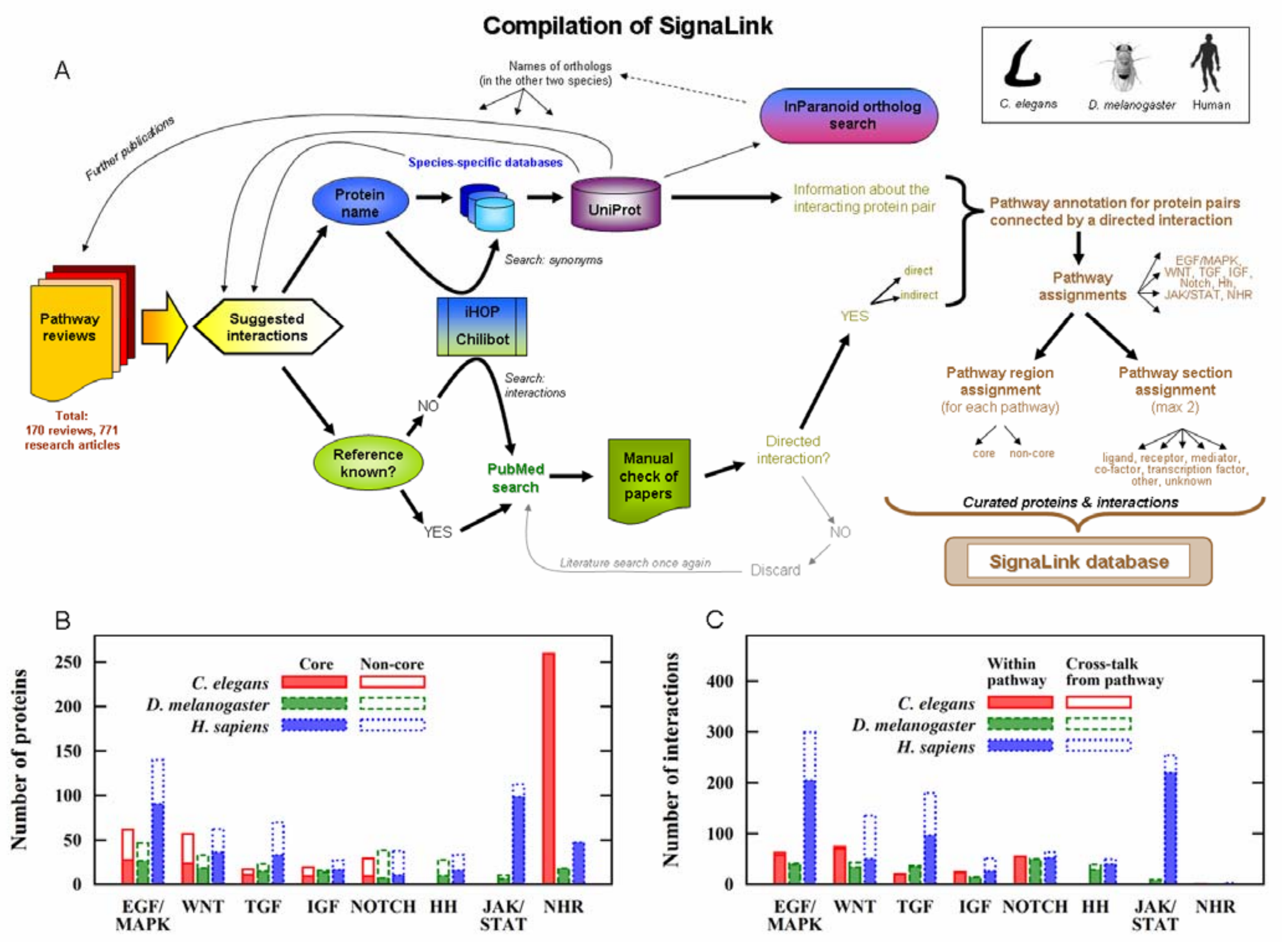

**Fig. 1.** Basic information about SignaLink. (A) The manual curation process. (B) Weighted protein numbers of SignaLink in the 8 signaling pathways of the 3 investigated species. (C) Weighted interaction numbers within pathways and between pathways (cross-talks). See the Supplement for details.

### 2.4 Functional annotation of drug target candidates

We collected information on the proteins that can be relevant in drug target discovery with DAVID (Dennis, Jr. et al., 2003). We downloaded disease-related annotations from OMIM, GAD, and Orthodisease (Amberger et al., 2009; Becker et al., 2004; O'Brien et al., 2004), domain information from InterPRO (Hunter et al., 2009), and molecular function and cellular component data from GO (Harris et al., 2004).

## 3 RESULTS

### 3.1 Uniform compilation of signaling pathways in three metazoan species

We curated the signaling pathways of the nematode *Caenorhabditis elegans*, the fruit fly *Drosophila melanogaster*, and *Homo sapiens*. From the wide variety of classification schemes for selecting signaling pathways (Bader et al., 2006) we followed the biochemical approach of Pires-daSilva and Sommer (Pires-daSilva and Sommer, 2003). We selected 8 major pathways for curation – EGF/MAPK, Ins/IGF, TGF-β, WNT, Hh (Hedgehog), JAK/STAT, Notch, and NHR (Nuclear Hormone Receptors) – that have central roles both in development and in normal cellular signaling.

SignaLink is a manually compiled resource integrating experimentally confirmed genetic and physical interactions from healthy tissue types. Proteins and interactions are listed without tissue-specificity and can be visualized as networks of potential interactions. Tissue- and disease-specific information can be added easily as shown in the examples below. Five combined characteristics create the unique utility of SignaLink:

(1) Pathways are biochemically defined and encompass all major developmental signaling mechanisms;

(2) A protein can belong to more than one pathway (if it does, then it is called a multi-pathway protein);

(3) Proteins are tagged with (i) the pathway(s), (ii) pathway region(s) (core, peripheral), and (iii) the pathway sections (one or two of: ligand, receptor, mediator, co-factor, transcription factor, other) they belong to;

(4) The level of detail is the same for the entire database;

(5) Interactions are directed and manually labeled with PubMed IDs (experimental evidence).

Currently, SignaLink lists 560 proteins and 237 interactions from *C. elegans*, 344 proteins and 233 interactions from *D. melanogaster*, and 646 proteins with 991 interactions from humans. Similarities and differences between species and pathways are shown in Figs. 1B and 1C. The database, its help pages, a detailed description of the curation process, and network visualizations of all pathways are available at http://SignaLink.org.

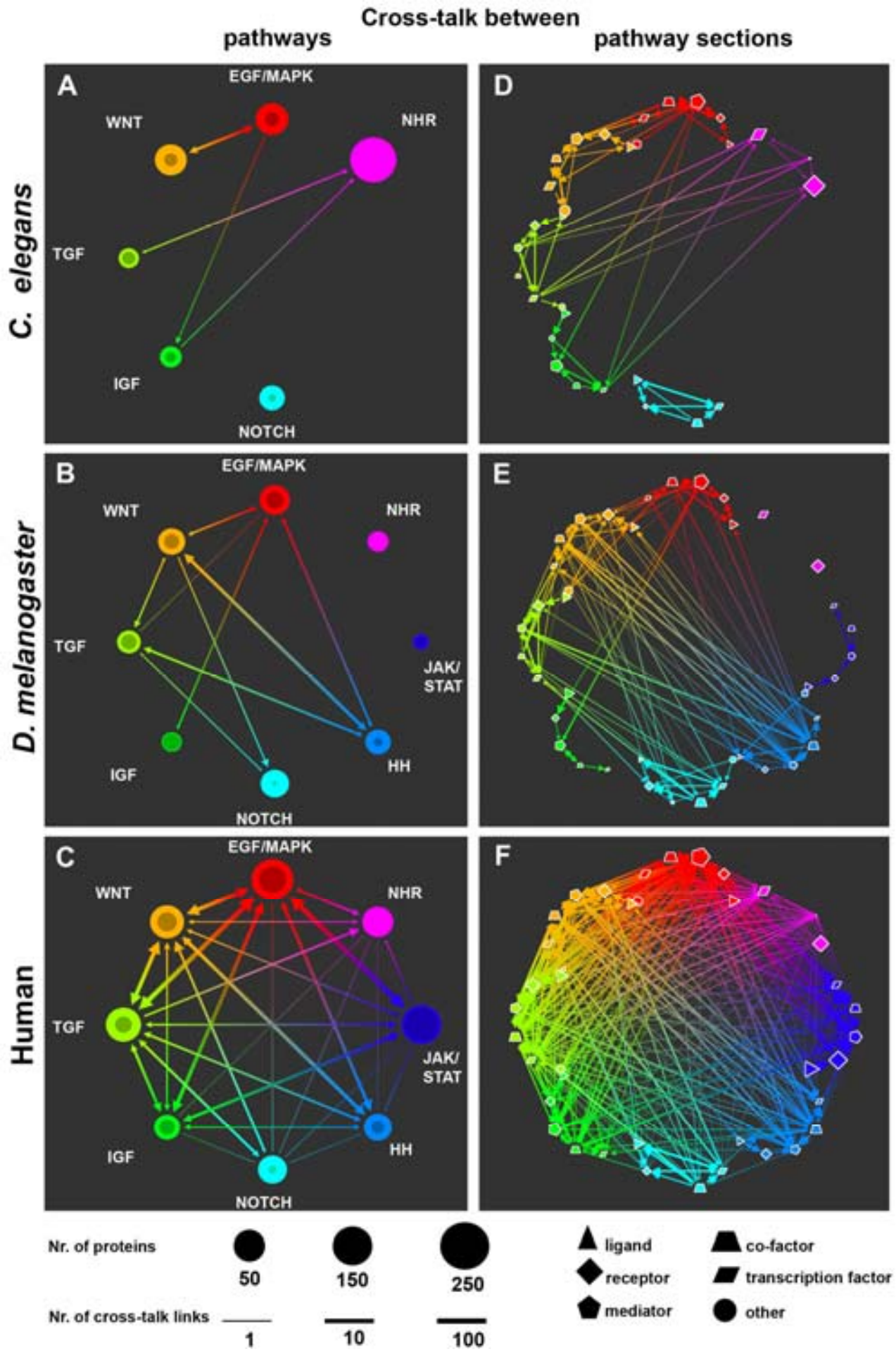

**Fig. 2.** Maps of signaling cross-talk in three metazoans integrating manually curated data from healthy tissue types. Pathway cores are indicated by darker colors. The width of a link between two pathways is proportional to the (weighted) number of directed signaling interactions (cross-talks) between the two pathways. A directed signaling interaction between two proteins annotated to $n$ and $m$ pathways, respectively, adds $1/(nm)$ to the weight between any two of the directed pathway pairs connected by this interaction. Two oppositely directed interactions are counted separately. (A-C) Cross-talks, *i.e.*, signaling interactions between pathways. Note that in humans any two pathways can cross-talk. (D-F) Cross-talk by pathway sections. In humans cross-talk is ubiquitous even among pathway sections, while in the other two species mostly co-factors and mediators cross-talk. See Methods for details and http://SignaLink.org for data sets.

## 3.2 A large-scale view of species-specific pathway and pathway section sizes

In all 3 organisms a few of the 8 pathways are central and abundant. Of all proteins 26% to 38% participate in the EGF/MAPK and WNT pathways, respectively. Other pathways with high protein numbers are NHR in the worm, Hh and Notch in the fly, and TGF and JAK/STAT in humans. Altogether in each species 68% to 85% of all signaling proteins participate in these pathways and 56% to 70% of all cross-talks involve the EGF/MAPK, TGF, or WNT pathways. *C. elegans* has almost identical numbers of core and peripheral proteins in each pathway (except for Notch and NHR), while in the other two species the ratio of core to peripheral proteins is around 1.5.

Pathway size differences between the 3 species are often related to the different environments to which the cells of these organisms have adapted. For example, ligands from the environment can easily reach the nuclei of the worm's cells, thus, the worm's NHR pathway is exceptionally large (58% of all signaling proteins). On the other hand, due to the large variety of signals that human cells are exposed to the human JAK/STAT pathway is oversized compared to the other two species (21% of all signaling proteins in humans vs. 0% and 4% in *C. elegans* and *Drosophila*, respectively).

In all 3 species EGF/MAPK and IGF have high numbers of mediators. However, environmental differences may affect pathway section sizes too. In *C. elegans* transcription factors – dominated by the NHR pathway – are the largest pathway section (39%). In the other two species co-factors by far outnumber other pathway sections (32% to 42%) and in humans JAK/STAT ligands and receptors are abundant.

## 3.3 Multi-pathway proteins: Proteins functioning in more than one signaling pathway

In *C. elegans*, *D. melanogaster*, and humans, we found 6, 12, and 62 multi-pathway proteins, respectively. Within one human signaling pathway the ratio of proteins functioning in at least one other pathway varies from 5% (Notch) to 46% (IGF). Interestingly, a single protein can be even a central (*i.e.,* core) component in more than one pathway. For example, the scaffold protein AXIN and the kinase GSK3 are both core components of more than one signaling pathway (Luo and Lin, 2004; Frame and Cohen, 2001).

We found that EGF/MAPK – the largest pathway – is the only one sharing proteins with all other pathways. On the other end of the spectrum are the Notch, JAK/STAT, and NHR pathways: their proteins are contained by 3 or 4 other pathways. These differences correlate well with the numbers of pathway functions. Note also that the set of 62 human multi-pathway proteins is enriched with disease-related proteins: 45% (28) of them are known to be disease-related, while in the 8 human signaling pathways only 25.5% (165 of 646) and among all human proteins listed by Ensembl only 20% (3,929 of 19,534). For both comparisons $p < 0.001$.

## 3.4 Cross-species comparison of cross-talks

Next, we focused on how the complexity of intracellular signaling increases with a growing complexity of the organisms. In *C. elegans* only 6 of the 8 curated pathways are active, and the Notch pathway is isolated (Fig. 2A). In addition, the cross-talk network of the pathways – where nodes represent pathways and links represent cross-talks – is sparse. Between the 6 active pathways only 5 of the 30 ( = 6×5 ) possible cross-talk types are present. In *Drosophila* all 8 curated pathways of SignaLink are active, but the NHR and JAK/STAT pathways are still isolated. Without these two pathways the cross-talk network is already significantly denser than in the worm: 16 of the total 30 possible cross-talk types are present (Fig. 2B). In humans – the most complex organism of the three – all 8 curated signaling pathways are active and almost all of the 56 possible cross-talk types are possible (Fig. 2C). The ubiquity of cross-talks (all 28 pathway pairs can cross-talk) expands both the repertoire of possible phenotypes and the system-level responses to environmental and pathological changes.

In *C. elegans* cross-talk is possible through receptors, mediators, and transcription factors (Fig. 2D). In the other two species all pathway sections can participate in cross-talk, except for the NHR and JAK/STAT pathways of *Drosophila*, where cross-talk occurs mostly at the transcriptional level and through mediators, respectively (Figs. 2E-F).

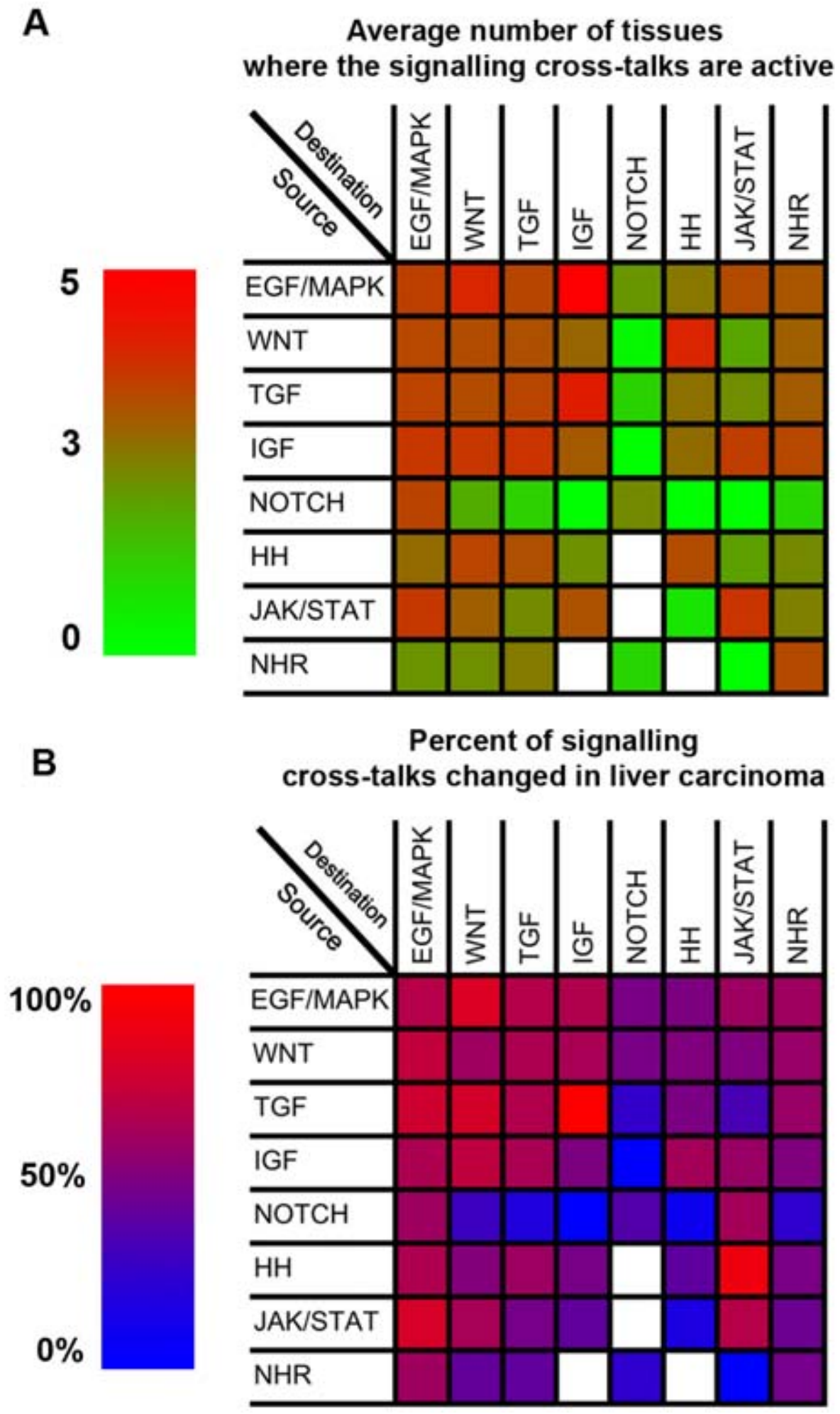

**Fig. 3.** Possible dynamical activity of the signaling interactions between and within pathways in (A) 5 selected healthy human tissue types and (B) 2 human liver carcinomas. (A) In the selected healthy tissue types the pathways EGF/MAPK, WNT, TGF, and IGF have the potential to cross-talk most frequently, while the cross-talk connections of the Notch pathway have less potential to be active. (B) In the two investigated liver carcinomas signaling interactions within the EGF/MAPK, WNT, and TGF pathways and their cross-talks are most extensively changed while the cross-talks of the Notch pathway are the least affected. See text for details.

In addition to the number of active pathways and cross-talks a further important indicator of signaling complexity is the number of cross-talks relative to all signaling interactions. In the worm 4.6% of all signaling interactions are cross-talks, in the fly 10.5%, and in humans 30.3%. Interestingly, the growth of the number of cross-talks from worm to fly and human is not simply due to the growth of the number of protein-coding genes (20 100, 13 800, 23 000, respectively) or the number of signaling-related PubMed articles (3 889, 11 367, 214 193 in worms, flies, and humans, respectively).

The presence of cross-talks in many pathways and pathway sections is a sign of the efficient utilization of resources: expanding the functions of an already existing pathway protein is more efficient than evolving a novel protein (Bhattacharyya et al., 2006). Given the high number of signaling cross-talks, a large variety of specific and robust phenotypes may emerge (Taniguchi et al., 2006). However, the actual signaling responses are controlled mainly by scaffold proteins, feedback loops, kinetic insulation, and the spatial and temporal expression patterns of proteins (Behar et al., 2007; Bhattacharyya et al., 2006; Kholodenko, 2006; Freeman, 2000). To map some of these possibilities, we investigated the dynamical activity of signaling cross-talks in humans where cross-talk was found ubiquitous.

## 3.5 Tissue- and disease-specific activity of cross-talks

Cross-talks, similar to other protein-protein interactions, are not active permanently in all tissue types. We considered an interaction to be possibly active in a given tissue type, if both of the mRNAs of its participating proteins are expressed in that tissue. It is reasonable to assume (as a simple approximation) that proteins whose mRNAs are expressed could be active in the given tissue (compared to those that are not transcribed). After merging SignaLink with protein expression profiles from human colorectal, muscle, skin, liver, and cardiovascular tissues, we quantified the possible tissue-specific activity of cross-talks. We found that cross-talks between the pathways EGF/MAPK and TGF, and cross-talks from WNT to TGF and from EGF/MAPK to JAK/STAT are overrepresented (Wald-test, upper limit: 7.25%) The complete statistical test is available in the Supplementary Data. In addition, we found that (i) the Notch and NHR pathways are the least connected to other pathways, and (ii) Notch and JAK/STAT are the least and most frequently used pathways, respectively: on average 26% and 48% of their proteins are expressed in the selected healthy tissue types (Fig. 3A).

Cancers are often viewed as systems diseases (Hornberg et al., 2006). In cancer cells large-scale modifications of signaling pathways, especially changes of cross-talks (Stelling et al., 2004), are prevalent. Accordingly, detecting which proteins (cross-talks) are differentially expressed (active) in a carcinoma tissue may point out key causes of the given tumour and can help the identification of novel, systems-based drug targets (Korcsmaros et al., 2007; Tortora et al., 2004). To do this we merged the network of 8 human signaling pathways with protein expression data from human liver carcinomas. We considered a signaling interaction to be altered in these two liver carcinomas, if, compared to healthy liver tissues, at least one of the participating proteins was differentially expressed (see Methods for details). In 3 of the 8 pathways (WNT, NHR, and JAK/STAT) only ~30% of all proteins were differentially expressed in the investigated liver carcinomas, while in the other 5 pathways this ratio was ~50% (Fig. 3B).

In summary, the largest signaling pathway, EGF/MAPK, is frequently used in the selected tissue types and significantly changed in the two liver carcinomas. JAK/STAT is also strongly used, but less modified in the two investigated liver carcinomas. A third example is Notch, which is neither strongly used nor strongly modified. Thus, even though cross-talks are possible between all pairs of investigated human signaling pathways, the possible activity of these pathways in healthy tissues and the modifications of their cross-talks in the analyzed cancer types are highly diverse. See the Supplementary Data for additional literature support.

### 3.6 Potential role of cross-talking proteins in drug target discovery

Signaling proteins are overrepresented among human disease genes (Sakharkar et al., 2007) and are intensively studied as potential drug target candidates (Chaudhuri and Chant, 2005), often with network-based methods (Berger and Iyengar, 2009). Among the 62 human multi-pathway proteins 21 (33.8%) are known drug targets compared to 15% (94 of 646) in all 8 pathways and 8.2% (1 610 of 19 534) among all human proteins. This implies that the remaining set of 41 human multi-pathway proteins may also be relevant for drug target discovery. In summary, the following two sets, altogether 327 proteins in our current study, are likely to be enriched with possible drug targets: (i) human multi-pathway proteins and (ii) proteins participating in cancer-related cross-talks.

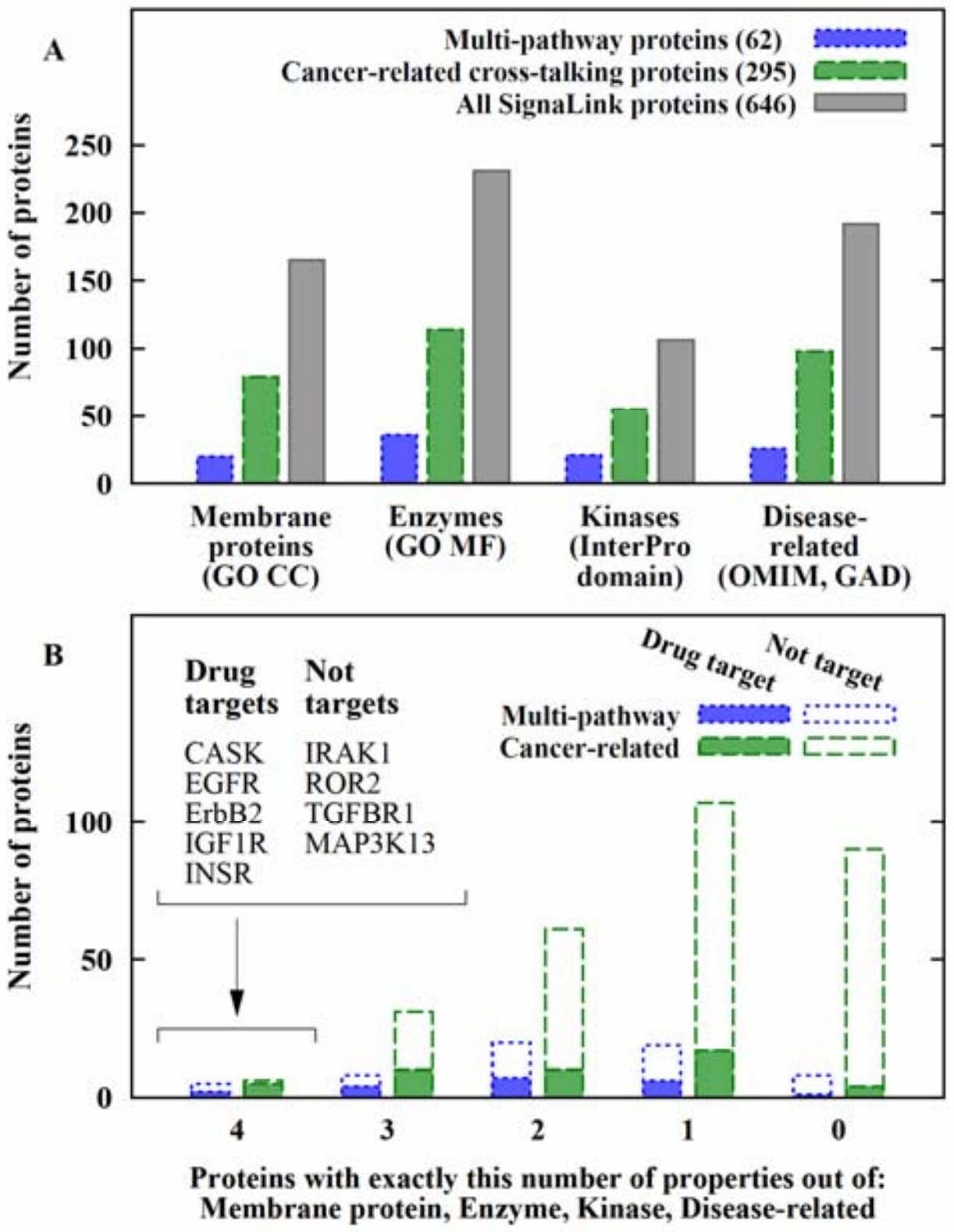

**Fig. 4.** Analysis of drug target candidate proteins listed with SignaLink. **A)** The numbers of membrane proteins (m), enzymes (e), proteins with kinase domains (k), and disease-related proteins (d) among the 3 groups of proteins listed in SignaLink as possible novel drug targets (control: all SignaLink proteins). **B)** Proteins with a fixed number of key properties out of the listed 4 (m, e, k, d). For each group the number of drug target proteins is shown separately. The names of the most promising candidates (all 4 key properties, not yet targeted) are listed.

We analyzed the drug target relevance of human signaling proteins by examining 4 key properties: disease-relatedness, localization in the plasma membrane (Gao et al., 2008; Yildirim et al., 2007), enzymatic functions, and kinase domain content (Fabbro et al., 2002). To identify the most promising drug target candidates from the 327 proteins selected in the previous paragraph, we first listed those 9 that have all 4 key properties, *i.e.*, disease-related enzymes (with a kinase domain) localized in the membrane (Fig. 4). Out of these 9 proteins 5 (CASK, EGFR, ErbB2, IGF1R, and INSR) are currently used as drug targets, while the other 4 (IRAK1, MAP3K13/LZK/MLK, ROR2, and TGFBR1) are not. ROR2 is essential during bone formation (Liu et al., 2007) and the other 3 proteins are inflammation-related factors (Klein et al., 2001). Interestingly, ROR2 was recently suggested as a therapeutic target for osteosarcoma based on expression analysis and siRNA treatment (Morioka et al., 2009). As for IRAK1, TGFBR1, and MLK, anti-inflammatory drugs (imiquimod, dexamethasone, and L-arginine, respectively) frequently affect their pathways, but without sufficient specificity (Klein et al., 2001). These findings support our predictions that these 4 human signaling proteins are promising novel drug targets.

## DISCUSSION

### 4.1 Advantages and limitations of SignaLink

According to a recent study (Cusick et al., 2009), manual curation projects (i) inherit the selection biases of the curated experiments, (ii) often lack the specific goals clearly defining the curation criteria, and (iii) it is usually difficult to estimate their completeness and reliability. Note, however, that many signaling proteins function outside the cell (ligands), in membrane-bound positions (receptors), or in the nucleus (transcription factors), and that proteins from these compartments are underrepresented in current high-throughput (HTP) data. Thus, for the purpose of identifying signaling proteins and interactions HTP techniques seem to be less adequate than those manual curation projects that have clear goals. In the case of SignaLink the precisely defined curation process combines original research articles and reviews, thus, both experimental evidence and its critical discussion by specialists are included.

Applying a biochemically-based, well-documented, and clear pathway definition is central to SignaLink. For example, the EGF/MAPK pathway in SignaLink contains (with evolutionary and biochemical reasoning) the pathway from the EGF ligand to the terminal MAPK kinases. In several other databases this pathway is scattered across many separate (sub)pathways (*e.g.,* EGFR, RAS, p38, JNK, ERK, ASK). An important consequence of precise pathway definitions is the reduced number of examined pathways. We suggest that appropriate and precise grouping, to avoid artificial pathway constructs, may be a better indicator of the goodness of the resource than merely the large number of pathways.

Of the constantly increasing number of signaling databases (Bader et al., 2006) many are proprietary, list fewer than 200 molecules, or only selected types of pathway components, *e.g.,* protein kinases. Even among the few databases passing these criteria (free for academic use, more than 200 molecules, all pathway component types included) there are currently no gold standards compiled with similar goals and methods as SignaLink. It is therefore important to compare both the curation protocols and the actual data of several available databases before selecting one of them for a particular analysis. In Table 2 we compare three widely used pathway databases – KEGG, Reactome, and NetPath – and SignaLink (see the Supplementary Data file for details). In each pairwise comparison we used the pathways available in both databases.

According to Table 2 and the Supplementary Data, SignaLink has the following advantages compared to the three analyzed databases: (1) precisely defined and documented curation protocol; (2) highest numbers of signaling proteins and interactions in the curated signaling pathways; (3) highest numbers of cross-talks and multi-pathway proteins; (4) largest protein overlap with the other databases; (5) above the average number of publications used per pathway; (6) minimal usage of protein isoform names; (7) no binary interactions inferred from the membership of two proteins in the same complex; (8) low number of proteins from UniProt/TrEMBL (*i.e.*, few unverified proteins). SignaLink was compiled based on pathway reviews and primary research articles. Thus, the high numbers of signaling proteins (including multi-pathway proteins) and cross-talks are likely to be dominated by true positives, indicating higher precision and coverage.

**Table 2.** Comparison of database content for human pathways between 3 manually curated databases and SignaLink. In each pairwise comparison we compared only the pathways curated in both databases.

| | **KEGG** | **Reactome** | **NetPath** | **SignaLink** |
|---|---|---|---|---|
| **Pathways** | MAPK, Insulin, Hedgehog, JAK/STAT, Notch, TGF-β,WNT | EGFR, Insulin receptor, Notch, TGF, WNT | EGFR1, Hedgehog, Notch, TGF, WNT | EGF/MAPK, IGF, Hh, JAK/STAT, Notch, NHR, TGF, WNT |
| **Number of proteins and interactions*** | | | | |
| **Proteins** | 429 (483) | 56# (348) | 362 (355) | 646‡ |
| **Interactions** | 1,502+ (990) | 682# (689) | 457 (701) | 991 |
| *cross-talks* | *225+ (228)* | *79# (226)* | *60 (171)* | *300×* |
| **Number of publications*** | 73 | 166 | 351 | 941 |

In KEGG only protein complexes and their interactions were available and we constructed a list of binary interactions, *i.e.*, a network, with the "matrix" method (Bader and Hogue, 2002). Symbols: * For the same pathways in SignaLink: EGFR, EGFR1, and MAPK are compared to the EGF/MAPK pathway of SignaLink; # binary interactions from membership in the same complex; ‡ including predicted pathway member proteins; + for each directed interaction between two protein groups (KEGG complexes) we added a directed link between each protein of the source group and each protein of the target group; ×weighted cross-talk interaction number: a directed signaling interaction between two proteins annotated to $n$ and $m$ pathways, respectively, adds $1/(nm)$ to the weight between any two of the directed pathway pairs connected by this interaction. See the Supplementary Data for further details.

Despite the care we have taken in creating SignaLink, it does have limitations, *e.g.,* SignaLink does not contain all signaling proteins. Only those signaling proteins have been included that have an experimentally verified function in the selected 8 major pathways or a directed interaction with at least one of their proteins. Several groups of proteins were fully excluded from SignaLink. These groups, together with our detailed reasons for excluding them, are listed in the Supplementary Data. The compilation of SignaLink was based on published review and research papers. Note that the curation of the current version of SignaLink was closed in May 2008. Naturally, more pathways (defined by the same evolutionary and biochemical rules) and proteins can be added in a future version. We plan to update SignaLink every July (starting 2011). The next update will include recent high-confidence high-throughput data. Overall, based on the comparison we presented in this sub-section, we expect that the limitations of SignaLink are small compared to the improvements it can provide.

## 4.2 Current applications

The primary goal of SignaLink is to provide *maps of global pathway communication* in 3 metazoans (*C. elegans*, *D. melanogaster*, and *H. sapiens*) with well-documented, uniform manual curation. Interactions from all healthy tissue types were included into SignaLink, and expression data from selected tissues were used for dynamic analysis. We found that the pathways EGF/MAPK, JAK/STAT, and Notch are clear examples for three distinct types of behavior: (i) high expression in normal tissue types and strong changes in cancer, (ii) high expression, but small changes, and (iii) low expression and small changes. See the Supplementary Data for additional literature support.

Network analyses – combined with system-level resources – can contribute to modern *drug target discovery*, *e.g.,* to polypharmacology and multi-target drug selection (Hopkins, 2008; Korcsmaros et al., 2007). With SignaLink one can prioritize novel drug target candidates by examining the list of (i) multi-pathway proteins and (ii) proteins participating in cancer-related cross-talks. Some of these proteins could be specific and proper targets, some of them could be too central and aspecific After listing the properties of these proteins relevant for drug target selection, we suggested 4 novel drug target candidates. One of them, ROR2, was recently proposed as a novel chemotherapeutic target, while the other 3 are known to be non-specifically affected by anti-inflammatory drugs. A broader list contains 35 additional proteins with a lower confidence, which may be again filtered with additional criteria, *e.g.*, phosphatase activity.

In biomedical experimental work the functions of a few selected proteins are often modified. These changes can unexpectedly perturb signaling pathways and non-specifically affect cellular processes. Based on several data sources, including SignaLink, we have launched a web server, PathwayLinker, to aid experimental work by linking the queried proteins to signaling pathways through physical and/or genetic interactions (Farkas et al., *submitted*).

## 4.3 Future applications

As their name suggests, targeting multi-pathway proteins may not be selective. However, selectivity is a key property in pharmacology, thus, analyses of multi-pathway proteins could support drug target discovery. Based on SignaLink we suggest single-gene knock-out experiments or RNA silencing of individual cross-talking proteins to help understand the *functional selectivity* of signaling pathways and to reduce the redundancies that are assumed to make many currently used drugs less efficient (Urban et al., 2007; Tortora et al., 2004).

For the *mathematical modeling* of the dynamical behavior of biomolecular pathways the precise, high-coverage reconstruction of the static network structure of these pathways is crucial (Papin et al., 2005). In the case of signaling pathways the manual curation of the existing literature can be efficient for assembling large-scale interaction maps. SignaLink data sets have a simple and uniform structure and are available in several formats for all 8 pathways and 3 species at http://SignaLink.org. This allows one to easily merge SignaLink with stoichiometric and expression data from, *e.g.*, perturbation analyses (Papin et al., 2005). The static network provided by SignaLink can serve as a backbone for both numerical and differential equation-based models and – due to SignaLink's focus on cross-talks – for deciphering the rules of cooperation between pathways (Wang et al., 2009; Borisov et al., 2009). Finally, perturbation analyses that cannot be carried out with HTP protein-protein interaction (PPI) networks (they list undirected PPIs), may be manageable with SignaLink, because it contains the directions of interactions.

A central goal of *synthetic biology* is to engineer cells that can carry out novel tasks (Friedland et al., 2009; Bhattacharyya et al., 2006). One way to achieve this goal is to rewire signaling circuits by modifying scaffold proteins and other biomolecules or by changing feed-back loops. In these studies detailed high-quality maps of intracellular signaling are essential, especially the positions of cross-talks and multi-pathway proteins.

*Comparative evolutionary studies* usually focus on conserved and altered mechanisms underlying the observed differences in body plans. Among metazoans these differences are largely due to changes in the complexity of regulation rather than different gene numbers (Levine and Tjian, 2003; Szathmary et al., 2001). Two decisive regulatory networks in this case are transcriptional control and signaling, in particular, cross-talks. SignaLink was compiled with uniform curation rules for 8 major pathways in 3 metazoans that have similar signaling mechanisms but different morphologies. Therefore, the data sets of SignaLink are well applicable to studying evolutionary changes.

### 4.4 Conclusions

Contrary to earlier views, signaling pathways are now understood as interlinked (not linear) routes heavily interlinked by cross-talks. This major paradigm shift necessitates novel, systems-based approaches, *e.g.*, new techniques for curation and modeling. Attempting to meet the novel curation requirements, we have compiled a signaling pathway resource, SignaLink. It allows the systematic comparison of pathways and their cross-talks. After finding that any two of the 8 selected human signaling pathways may cross-talk, we quantified the possible activity patterns of these cross-talks in healthy and cancerous human tissues. Large-scale mapping of pathways and the activity patterns of their cross-talks revealed multi-pathway and cancer-related cross-talking proteins that can be relevant for drug target discovery.

## ACKNOWLEDGEMENTS

We thank B. Papp, F. Jordan, and L. Zsakai for comments; P. Connolly for proofreading; T. Vicsek for discussions; G. Szuromi, R. Palotai, and P. Pollner for technical help. The authors are grateful to the anonymous referees for their comments and suggestions.

*Funding*: This work was supported by the EU 6th Framework Programme (FP6-518230), the Hungarian National Science Fund (OTKA K69105, K75334, K68669), and the Hungarian National Office for Research and Technology (CellCom RET). IJF is supported by an EEA Fellowship.

**Supplementary Material for**

# Uniformly curated signaling pathways reveal tissue-specific cross-talks and drug target candidates

Tamás Korcsmáros[1,2,*], Illés J. Farkas[3,*], Máté Szalay[2], Petra Rovó[1], Dávid Fazekas[1], Zoltán Spiró[2], Csaba Böde[4], Katalin Lenti[5], Tibor Vellai[1], and Péter Csermely[2,+]

[1] Department of Genetics, Eötvös University, Pázmány P. s. 1C, H-1117 Budapest, Hungary
[2] Department of Medical Chemistry, Semmelweis University, PO Box 260, H-1444 Budapest, Hungary
[3] Statistical and Biological Physics Research Group of the Hungarian Academy of Sciences, Pázmány P. s. 1A, H-1117 Budapest, Hungary
[4] Morgan Stanley Hungary Analytics Ltd, Lechner Ö. f. 8, H-1095 Budapest, Hungary
[5] Department of Morphology and Physiology, Semmelweis University, Vas u. 17, H-1088 Budapest, Hungary
* equal contributions
[+] Corresponding author, csermely@eok.sote.hu

## See the supporting website for further information: http://SignaLink.org

## Contents

# Supplementary results and data

## *Comparison of pathways across species*

On the system level one may observe that the worm has more signaling proteins than the fly, which is mostly due to the large number (more than 250) of nuclear hormone receptors (NHR) in the worm. As for single pathways, note that in the worm several proteins orthologous to fly and human JAK/STAT and Hh pathway member proteins are expressed. However, only few interactions of these proteins have been experimentally verified so far (Burglin and Kuwabara, 2006). In other words, experimental evidence cannot support yet that in *C. elegans* these two pathways are functional. It has been suggested that the pathways JAK/STAT and Hh disappeared from the worm during its evolution (Pires-daSilva and Sommer, 2003). A comparison of pathways between the three species shows also that in all three the largest pathway (containing the largest number of proteins) is the EGF/MAPK pathway, while the next three (in the order of protein number) are WNT, TGF, and IGF, respectively. In *C. elegans* and *D. melanogaster* the total weighted number of proteins[1] in these four pathways is approximately the same (154.5 and 117.3 proteins, respectively), while in humans the total is approximately twice this number (295.2). As a contrast, the Hh and Notch pathways have very similar protein numbers in all three species. Note that in humans four (EGF/MAPK, IGF, JAK/STAT, and WNT) of the five largest pathways are kinase-based pathways.

A look at the internal structure of the eight curated signaling pathways reveals further details. See the accompanying website, http://SignaLink.org, for detailed statistics, a detailed curation manual, and zoomable images of the 8 curated pathways in all 3 organisms. In humans the backbone (core region[2]) of the EGF/MAPK pathway is significantly stronger due to the high number of parallel MAPK cascades. Similarly, the backbone of the WNT pathway is notably larger in humans than in the other two species, which is caused mainly by the high numbers of receptors and ligands. As a contrast, in the IGF, Hh, Notch, and TGF pathways the ratios between the sizes of pathway regions (core and periphery) are almost identical in the three species (in the Notch pathway the core region is small). The JAK/STAT pathway changes strongly across the three investigated species. In *C. elegans* the pathway is not present, in *Drosophila* it has a few proteins (a similar number of core and peripheral proteins) while in humans it is already comparable in size to the largest pathways and its core region is larger than its peripheral region. The large number of human JAK/STAT pathway member proteins and the strong core region of this pathway are mainly due to the high number of ligands (cytokines) and their receptors allowing this pathway to produce a combinatorially large repertoire of immune responses. In other words, the change of the size and structure of the JAK/STAT pathway reflects a major functional change: acquiring a central role in immune responses.

Because of the large number of its experimentally known genetic interactions, the fly has currently many more known interactions among its signaling proteins than the worm. However, experimental approaches have not yet verified a large portion of the genetic interactions in the fly as true physical interactions, and so the evidence provided by genetic interactions is less direct than other evidence types. Thus, before inserting genetic interactions into SignaLink, we have carefully filtered them, mostly by searching for other types of evidence available for the same interactions. See below under “Manual curation process of SignaLink” for a description for interaction evidence types. As a result in SignaLink the signaling interactions of the fly and the worm have similar levels of confidence and their total numbers are close too.

---

[1] A protein participating in *n* pathways adds 1/*n* to the weighted number of proteins in each of these pathways.
[2] For an explanation of the term "core" see below in this document under "Manual curation process of SignaLink".

As the fly is a more complex organism than the worm, one may expect that the number of high-confidence protein-protein interactions in it, including directed signaling interactions, is larger than in the worm. Consequently, we expect that many of the genetic interactions in the fly will be experimentally verified in the near future as true physical protein-protein interactions, many of them as cross-talks (*i.e.*, signaling interactions between pathways).

### Statistical significance tests of cross-talk numbers

We have determined the adjusted Wald Confidence Interval to find the pairs of pathways between which pathway-pathway connections (*i.e.*, cross-talks) are more frequent than the others (Agresti and Coull, 1998). With the adjusted Wald confidence interval we determine the 95% confidence interval, supposing binomial distribution of the cross-talks. If the number of cross-talk lies out of this 95% confidence interval, we may say that the particular pathway connections are more frequent. Supplementary Table 1 shows the relative frequencies of the interaction numbers between pairs of human signaling pathways in SignaLink. Note that this table shows only the 10 highest scoring directed pairs of pathways. Based on the full list of 56 ( = 8 * 7 ) directed pairs of pathways the average of the relative frequency is is 1.79% the upper border of the adjusted Wald confidence interval lies at 7.27%.

<table>
<tr><th colspan="2">Directed cross-talk between pathways</th><th>Number of interactions</th><th>Relative frequency</th><th></th></tr>
<tr><td>IGF</td><td>EGF/MAPK</td><td>36</td><td>3.10%</td><td></td></tr>
<tr><td>EGF/MAPK</td><td>IGF</td><td>50</td><td>4.31%</td><td></td></tr>
<tr><td>EGF/MAPK</td><td>WNT</td><td>53</td><td>4.57%</td><td></td></tr>
<tr><td>TGF</td><td>WNT</td><td>57</td><td>4.91%</td><td></td></tr>
<tr><td>WNT</td><td>EGF/MAPK</td><td>65</td><td>5.60%</td><td></td></tr>
<tr><td>JAK/STAT</td><td>EGF/MAPK</td><td>83</td><td>7.16%</td><td></td></tr>
<tr><td>EGF/MAPK</td><td>JAK/STAT</td><td>90</td><td>7.76%</td><td rowspan="4">Statistically significantly overrepresented</td></tr>
<tr><td>WNT</td><td>TGF</td><td>90</td><td>7.76%</td></tr>
<tr><td>TGF</td><td>EGF/MAPK</td><td>104</td><td>8.97%</td></tr>
<tr><td>EGF/MAPK</td><td>TGF</td><td>160</td><td>13.79%</td></tr>
</table>

**Supplementary Table 1.** Wald statistical scores of the interaction numbers between the 8 human signaling pathways in the SignaLink database. Only the 10 directed pathway pairs with the largest scores are shown here. Relative frequencies above 7.25% are statistically significant (see text above for details).

## *Analysis of cross-talks in each species*

In this section we first define *cross-talks.* Then we describe and discuss active pathways in the three organisms. Last, we show examples for cross-talking pairs of pathways.

The *classical definition* of cross-talk assumes that each protein is contained by exactly one signaling pathway. According to this definition, if proteins A and B belong to different signaling pathways, then the signaling *interaction* between proteins A and B is a *cross-talk between their signaling pathways*. SignaLink – and several other databases too – provide a more precise mapping of signaling pathways and allow that a signaling protein belongs to more than one pathway. Therefore, the classical definition of cross-talk has to be extended. A signaling interaction between two proteins, A (contained by pathways $a_1$, $a_2$, ...) and B ($b_1$, $b_2$, ...), is at the same time a cross-talk between each pair of different a-b pathways, *i.e.*, between any pathway pair $a_i$-$b_j$, supposed that these two pathways are different. Of course, the signaling interaction between proteins A and B is a signaling interaction within each such $a_i$ pathway that is listed among the $b_j$ pathways as well.

As an example consider a signaling network containing a total of four signaling proteins – A, B, C, and D – and two directed signaling interactions: A→B and C→D. Assume that in this example network protein A is contained by the pathways EGF/MAPK and Hh, protein B is contained by the pathways EGF/MAPK and WNT, protein C is contained only by the IGF pathway, and protein D is contained only by the Notch pathway. The directed signaling interaction between proteins A and B is at the same time a (directed) cross-talk between the pathways EGF/MAPK and WNT, a (directed) cross-talk between the pathways Hh and EGF/MAPK, a (directed) cross-talk between the Hh and WNT pathways, and it is also a (directed) signaling interaction inside the EGF/MAPK pathway. The signaling interaction between proteins C and D is a (directed) cross-talk between the IGF and Notch pathways.

Note that if a unit weight is assigned to each of these five (directed) pathway-pathway interactions (one cross-talk between each of the three directed pathway pairs EGF/MAPK→WNT, Hh→EGF/MAPK, and Hh→WNT, and one directed interaction inside the EGF/MAPK pathway due to the interacting (directed) protein pair A→B; one cross-talk between the directed pathway pair IGF→Notch due to the interacting (directed) protein pair C→D), then the total weight of the signaling interaction of the (directed) protein pair A→B would be 4 compared to the total weight of the signaling interaction of the (directed) protein pair C→D, which would be only 1. This would incorrectly overestimate the biological significance of the first pair (A→B) compared to that of the second pair (C→D). For most interacting pairs of signaling proteins there is no or little *quantitative* information about the total (system-level) effect of their signaling interaction. Thus – due to the absence of such information – we propose that each signaling protein-protein interaction should add the same total weight, 1, to pathway-pathway interactions. Note that this total weight includes both cross-talks ($a_i$ and $b_j$, the selected pathways of the two interacting proteins, are different pathways) and interactions within pathways ($a_i$ and $b_j$ are the same pathway).

If in a signaling interaction at least one of the two interacting proteins belongs to more than one pathway, then usually there is no or little *quantitative* information about which one of the connected pathway pairs is linked more strongly by this pair of interacting proteins. Therefore, we propose that if a protein belongs to $n$ pathways and another protein that it interacts with through a directed signaling interaction belongs to $m$ pathways, then the directed signaling interaction between these two proteins should add the same weight, $1 / (n * m)$, to the cross-talk weight between any two of the directed pathway pairs connected by this directed signaling interaction. In the example network (defined above) the directed signaling interaction of proteins A and B adds a cross-talk weight of 1/4 to the three (directed) cross-talks EGF/MAPK→WNT, Hh→EGF/MAPK, WNT→TGF, and the interaction of the EGF/MAPK pathway with itself, while the directed signaling interaction of proteins C and D adds a cross-talk weight 1 to the cross-talking (directed) IGF→Notch pathway pair.

Note that since protein-protein interactions are directed, two oppositely directed interactions between the same proteins, *i.e.*, A→B and B→A, are counted separately and contribute separately to the interaction weights (cross-talks and within-pathway) between the directed pairs of their pathways.

## Active pathways and cross-talking pathways

With an increasing complexity of the investigated organisms the number of active pathways increases and the number of cross-talking pathways also increases. In other words, the cross-talk map of the organism becomes denser.

In *C. elegans* only six of the eight curated pathways are active. In addition, the network of pathways is sparse: only 5 of the 30 possible cross-talk types (connecting the 6 active pathways) are present (see the supporting website for detailed statistics). Interestingly, the number of cross-talks in the two possible directions between the same pair of pathways is often different. For example, in *C. elegans* – according to SignaLink – the weighted number of directed signaling interactions pointing from the EGF/MAPK pathway to the IGF pathway is 1.5, while in the opposite direction (IGF→EGF/MAPK) there is no signaling interaction. Two further examples are (i) the cross-talks between the Hh and WNT pathways in *D. melanogaster* (for Hh→WNT the weighted number of directed interactions is 6.9 and for WNT→Hh it is 3.1) and (ii) the human EGF/MAPK and TGF pathways where the weighted number of directed signaling interactions pointing towards the EGF/MAPK pathway is much higher as in the opposite direction (47.2 vs. 28.0).
In *Drosophila* all eight curated pathways of SignaLink are active (see Fig. 2b of the main text), however, the NHR and JAK/STAT pathways are isolated. Without these two pathways the cross-talk network is already relatively dense: 16 of the total 30 possible cross-talk types are present.

In humans all eight curated signaling pathways are active and all 28 ( = 8 * 7 / 2 ) pathway pairs are connected by cross-talks. As signaling interactions are directed, the 28 pairs of pathways allow 56 ( = 8 * 7 ) types of directed cross-talk. Of these 56 only 6 are inactive. The highest (weighted) number of cross-talks is between EGF/MAPK and the other four large pathways: WNT, TGF, IGF, and JAK/STAT. In contrast to the other two species, the NHR pathway is active and cross-talks with all other pathways, which is mainly due to the signaling interactions between Nuclear Hormone Receptor proteins and the transcription factors of the other pathways.

## Examples: Cross-talks in healthy human tissues

To underline the importance of signaling cross-talk in the three investigated species, we list here first a few general examples about the role of cross-talk in organ development. Then we provide examples for cross-talk between two core proteins, a core and a peripheral protein, and two peripheral proteins.

In *C. elegans* EGF/MAPK and WNT signaling can cooperate to establish embryonic polarity and to specify cell fate. The decision between reproductive and dauer development is jointly controlled by TGF and IGF signaling, while vulval development is determined by the cross-talks between EGF/MAPK, WNT, and Notch signaling (Hanna-Rose and Han, 2000). The patterning of the *Drosophila* retina is cooperatively controlled by multiple pathways: Hh, Decapentaplegic (Dpp) from the TGF-beta family, Notch, and EGF/MAPK. Similarly, the development of the wing is controlled jointly by Hh, Dpp, WNT, and EGF/MAPK (Baker, 2007). Recent studies show that the development of human hematopoietic stem cells is jointly controlled by the Hh, JAK/STAT, Notch, TGF, and WNT pathways (Blank et al., 2008).

Interestingly, a cross-talking pathway may interact through not just one, but many of its pathway sections (receptor, mediator, *etc.*) and both pathway regions (core, periphery). In humans ALK-6 and ROR2 are core proteins of the TGF and WNT pathways, respectively, and they are both receptors (Sammar et al., 2004). Similarly, EGF/MAPK and IGF cross-talk through their core proteins GRB2 and IRS1 (Claeys et al., 2002), two receptor adaptor proteins, while EGF/MAPK, Hh, and WNT cross-talk through their core proteins RSK1 and GSK3 (De Mesquita et al., 2001), which are mediators (kinases). A cross-talk involving a core and a peripheral protein is the interaction between PIAS4 (JAK/STAT) and SMAD4 (TGF) (Long et al., 2003), and a cross-talk between two non-core proteins is the interaction of CTBP1 (Notch) and HIPK2 (TGF) (Zhang et al., 2003).

## *Multi-pathway proteins (proteins contained by more than one pathway)*

### Protein numbers and examples

During the curation process we found that several proteins are contained by more than one signaling pathway. We call these proteins here multi-pathway proteins. We note that multi-pathway proteins and their possible functions were reported in (Komarova et al, 2005). In the worm, the fly, and humans we found a total of 6, 12, and 62 such proteins, respectively. In humans the eight curated pathways in SignaLink have the following numbers of multi-pathway proteins (*i.e.*, proteins that are contained by at least one other pathway too): EGF/MAPK (41), Hh (9), IGF (17), JAK/STAT (16), NHR (3), Notch (2), TGF (29), and WNT (22). In summary, within one human signaling pathway the ratio of multi-pathway proteins varies from 5% (Notch) to 46% (IGF).

Supplementary Figure 1 shows the eight human signaling pathways curated in SignaLink and the numbers of multi-pathway proteins. Observe that EGF/MAPK – the pathway containing the largest number of proteins – is the only pathway that shares proteins with all other pathways. On the other end of the spectrum are the Notch, JAK/STAT, and NHR pathways; they share proteins only with 3 or 4 other pathways.

We list now a few examples for human multi-pathway proteins. First, the scaffold protein Axin and the kinase GSK3 are both central components, *i.e.*, core proteins, in more than one signaling pathway (Luo and Lin, 2004; Frame and Cohen, 2001). Second, the WNT and Hh pathways both contain the same essential ubiquitin ligase complex (Maniatis, 1999). Moreover, ubiquitin ligases highly similar to this complex tune the TGF and Notch pathways (Itoh and ten Dijke, 2007; Itoh et al., 2003).

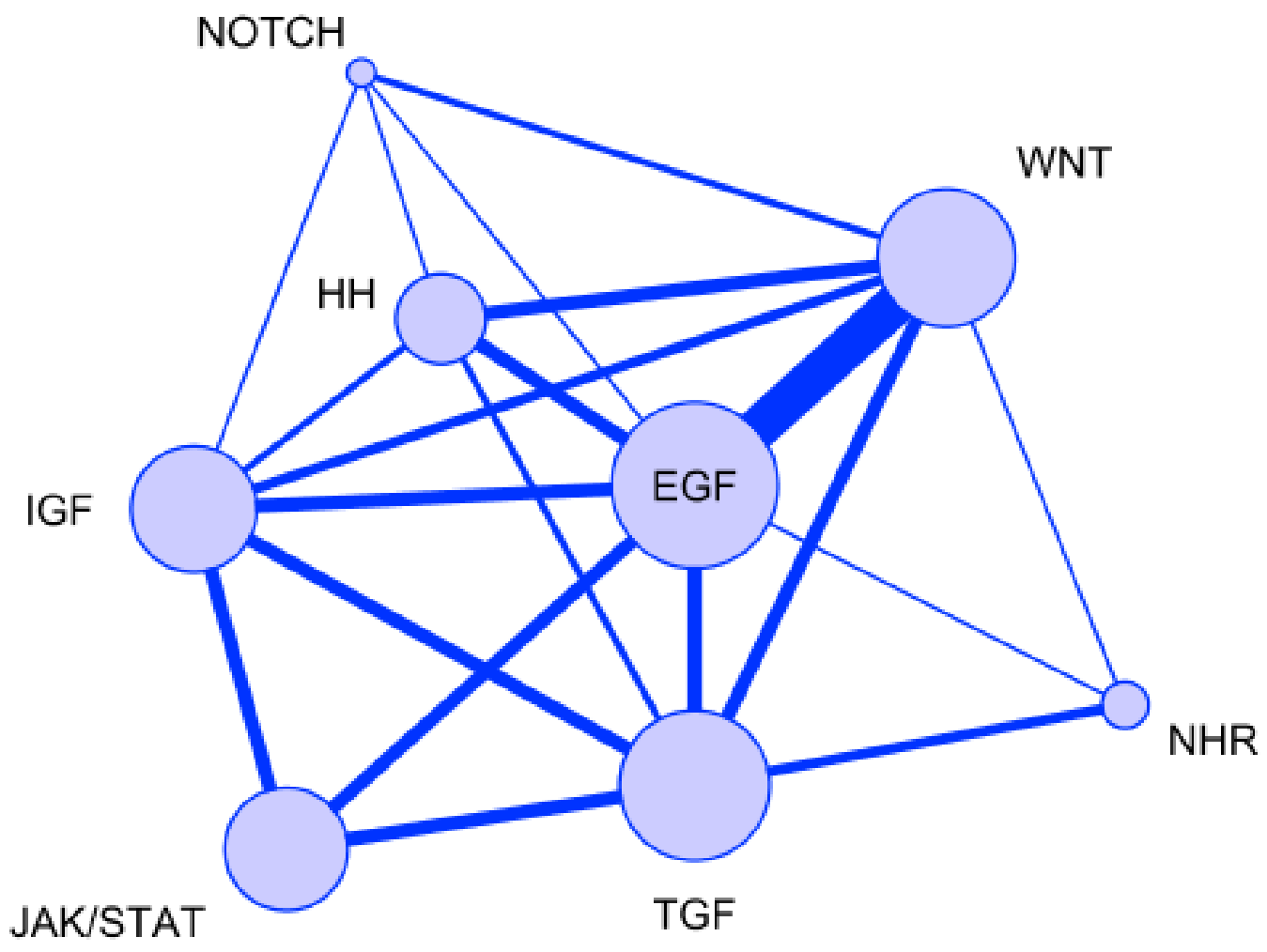

**Supplementary Figure 1.** In the 8 human signaling pathways curated in SignaLink there are 62 multi-pathway proteins. In this figure each node represents one human signaling pathway. The size of a node is proportional to the number of proteins contained by the pathway (represented by the node) and the width of a link between two nodes is proportional to the number of proteins contained by both pathways. See the text for details.

## Statistical significance tests: Disease-related proteins among multi-pathway proteins

Khi-square tests showed the significance of known disease-related proteins (Amberger et al., 2009; Becker et al., 2004; Dennis, Jr. et al., 2003) among multi-pathway proteins in different sets of proteins: The set of 62 human multi-pathway proteins is enriched with disease-related proteins: 45% (28) of them are known to be disease-related, while in the 8 human signaling pathways only 25.5% (165 of 646): $\chi^2$ =10.9817; p= 0.0009. The ratio of multi-pathway proteins among all human proteins listed by Ensembl is only 20% (3,929 of 19,534): $\chi^2$ = 24.0607; p= $0.9*10^{-6}$.

## *Cross-talks in the selected healthy tissue types and liver carcinomas*

According to eGenetics (Kelso et al., 2003) in the investigated healthy human tissues 40% (muscle) to 52% (colorectal) of the signaling proteins listed by SignaLink are expressed. We obtained protein expression in two screens of liver carcinomas (Wurmbach et al., 2007; Chen et al., 2002) from Oncomine (Rhodes et al., 2007). In these two liver carcinomas we considered a protein differentially expressed if the p value of its expression in at least one of the two screens – compared to healthy liver tissues and computed by a t-test of Oncomine – was below 0.05. We found that 46% of all SignaLink proteins have a modified expression compared to healthy liver tissue. Despite the similar ratios of active signaling proteins and the possibility of cross-talk between any two pathways, cross-talks are active in a tissue-specific manner. This flexibility allows the development and maintenance of different phenotypes, even in extreme cases such as cancer.

We continue with two examples for the tissue-specific expression of cross-talk types. First, observe in Fig. 3a of the main text that (in the selected tissue types) between the following pairs of large pathways the ratio of active signaling cross-talks is lower than the average: EGF/MAPK-IGF, IGF-JAK/STAT, EGF/MAPK-JAK/STAT, IGF-TGF, and IGF-WNT. Taking the pathway pair EGF/MAPK-IGF as an example, we found only two primary publications reporting signaling cross-talk between them (Roudabush et al., 2000; Vardy et al., 1995). The second sexample, standing in contrast to the first, is the following. For 3 larger (and several smaller) pairs of pathways cross-talk is more frequent in the investigated tissue types than one would assume from the sizes of the two cross-talking pathways: EGF/MAPK-NHR, NHR-TGF, and NHR-WNT. These fields are colored red in the chart on Fig. 3a. One reason for the higher than average ratio of active cross-talks between these four pathways is that SMAD3, a likely key protein in fibrogenesis (Flanders, 2004; Roberts et al., 2001), participates in all four of them and it is expressed more often in the selected tissue types than the average signaling protein.

Next, we discuss the figure on liver carcinomas (Figure 3b) of the main text. In general liver carcinomas are known to involve genetic mutations in the WNT and JAK/STAT pathways (Takigawa and Brown, 2008; Niwa et al., 2005), an increased activity of the EGF/MAPK and IGF pathways and their cross-talk (Desbois-Mouthon et al., 2006), and also an increased activity of the JAK/STAT pathway (Calvisi et al., 2006). A less known result is that several proteins of the Hh pathway are activated in human HCCs (hepatocellular carcinomas) (Huang et al., 2006). In addition to these Figure 3b of the main text shows the amount by which each cross-talk type is related to the investigated liver carcinomas. The cross-talks of EGF/MAPK with most other pathways are frequently changed (turned on or off), similarly to the interactions within the group EGF/MAPK-WNT-TGF-IGF. On the other hand, the cross-talk activity of NHR and Notch changes only weakly in the analyzed human liver carcinomas. (The low number of cross-talks of the Notch pathway may distort the results for this pathway.)

Finally, we concluded that the pathways EGF/MAPK, JAK/STAT, and Notch are clear examples for three distinct types of signaling behavior: (i) high expression in normal tissue types and strong changes in cancer, (ii) high expression, but small changes, and (iii) low expression with small changes. Below, we show literature support for these 3 major claims.

The high expression of the EGF/MAPK pathway in all selected tissue types corresponds to its central role in the evolution and normal function of complex signaling networks (Manning et al., 2002). Moreover, deregulation of its pathway members is the major cause of several cancer types (Amit et al., 2007). The JAK/STAT pathway is the main pathway utilized by immunologically important receptors, *e.g.*, cytokine receptors. These receptors are present in many tissue types, and similar to other highly expressed JAK/STAT pathway members they receive and transmit signals from immune cells to the given cell's nucleus (Haan et al., 2006). The Notch pathway is an important pathway for cell differentiation but less important in mature cell types, therefore most of its pathway members – many that are responsible only for the regulation of the pathway – are not expressed in the selected tissue types (Bolos et al., 2007). However, the Notch pathway can be crucial in the development of liver cancer: a recent study showed that in normal cells it could inhibit other oncogenic pathways (*e.g.*, WNT) but in liver cancer major components of the Notch pathway is downregulated (Wang et al., 2009b).

## *The website of SignaLink: Search, Browse, and Download options*

At http://signalink.org the user can **search** for signaling proteins and interactions of *C. elegans*, *D. melanogaster*, and humans by protein name, UniProt ID (or accession), and species-specific IDs (WormBase ID, FlyBase ID, and Ensembl protein ID). Result pages contain the following information (all hyperlinked): protein IDs (protein name, UniProt ID, UniProt accession, and species-specific IDs), the roles of proteins (pathway sections: ligand, receptor, mediator, co-factor, transcription factor, and other; and pathway regions: core and peripheral), orthologous proteins, and interaction types (incoming, outgoing, within pathways, and between pathways, *i.e.*, cross-talk).

The user can **browse** the pathways in each species and view zoomable high-resolution maps (networks) of each pathway colored by (i) pathway sections or (ii) by pathway regions or (iii) by ortholog information. For pathway sections and regions networks were exported from Cytoscape 2.5.2 using the Cytoscape plugins GOlorize (Release date: Feb. 12, 2007) and BiNoM (1.0). Node colors are listed on the website. A pie chart (with GOlorize) indicates that the given protein belongs to more than one pathway or more than one pathway section. We used the following parameters for the force-directed layout algorithm producing the networks. Default spring coefficient and length: both -10, node mass: 6, number of iterations: 1000, selected integration algorithm: Runge-Kutta (minimum and maximum edge weights: 0 and a very large number, respectively).

With the **ortholog network viewer** the user can explore all 8 curated signaling pathways separetely or together in each of the 3 species. This viewer shows the predicted pathway member proteins too. In all visualizations direct and indirect interactions are shown as normal and dashed lines, respectively. Arrows and blunted arrows stand for stimulation and inhibition, respectively.

You can **download** from http://signalink.org the curated signaling pathways of each species separately, all pathways of one selected species, and all pathways of all three species. Available formats are the following. (i) *All interactions* in a simple Excel file; (ii) *signaling proteins and interactions* in these formats: MS Excel, CSV (comma separated values), SQL dump, SBML (Systems Biology Markup Language), CYS and SVG files exported from Cytoscape, and PNG formatted images; (iii) *multi-pathway proteins* listed in a plain text file with TAB-separated fields; (iv) *predicted pathway member proteins* in an MS Excel file; (v) the *list of review papers* used for compiling SignaLink, grouped by pathways, in RefMan, MEDLINE, and BibTeX formats; (vi) the list of most likely *drug target candidates,* sorted by the number of analyzed properties that make them possible drug targets. Note that in the downloadable Cytoscape CYS files proteins functioning in more than one signaling pathway are colored pink, because currently the parameters of GOlorize cannot yet be fully exported into CYS files.

# Supplementary methods

## *Detailed description of the SignaLink database*

SignaLink uses 8 biochemically defined pathways important in metazoans: EGF/MAPK, Insulin and Insulin-like Growth Factor (IGF), TGF-β (TGF), Wingless (WNT), Hedgehog (Hh), JAK/STAT, Notch, and Nuclear hormone receptors (NHR). Instead of focusing on one particular function or phenotype, SignaLink defines each pathway by its unique combination of biochemical characteristics. For example, the JAK/STAT pathway has cytokine receptors, it is dominated by non-receptor tyrosine kinases, and its proteins frequently dimerise. Proteins are assigned to pathways (i) based on reviews of pathways and protein families and (ii) using biochemical interactions between proteins which are manually identified from the original publications reporting experimental evidence.

For each protein SignaLink stores four identifiers: (i) the UniProt ID of the protein (*e.g.*, P92172), (ii) the name of the gene (*e.g.*, *daf-7*), (iii) a species-specific gene ID (for *C. elegans* and *D. melanogaster*) or the protein's Ensembl ID (for humans), and (iv) an internal database ID (a positive integer). In addition to these each protein in SignaLink is manually tagged with (i) the pathway(s) it participates in (one or more of the 8 pathways curated in SignaLink), (ii) its pathway region(s) (core, peripheral) in each of its pathways separately and (iii) its pathway section(s) (one or two of: ligand, receptor, mediator, co-factor, transcription factor, and other).

In SignaLink we allowed a maximum of two pathway sections for each protein. A few specific cases and consequences are the following. First, we note that, strictly speaking, in the Notch pathway most proteins have more than two roles: receptor, mediator, and transcription factor. However, we labelled Notch proteins only with the pathway sections receptor and transcription factor, and not with mediator, because when Notch proteins are mediating they have almost no signaling interactions. Second, in the Nuclear hormone receptor pathway we listed only the actual NHR proteins (each both as receptor and transcription factor) and considered each NHR signaling protein to belong to the core. No ligands were included, because in the NHR pathway ligands are not proteins, but lipids. We also skipped the co-factors of the NHR pathway, because their function is nuclear regulation and not the adjustment of the signaling flow. We believe that the co-factors of nuclear hormone receptors and other nuclear factors, *e.g.*, chromatin remodelling factors, should be examined in the transcription regulatory network, not in the network of signaling interactions, thus, they belong to another type of cellular database.

The pathway region of a protein indicates whether it is essential in the signaling flow of the given pathway. Based on experimental evidence we allowed during the curation process explicitly that a protein belongs to more than one pathway and even (this happened rarely) that it is both a core and a peripheral component within the same pathway. As an extreme case the human protein SMAD3 belongs to four pathways and five regions within those four pathways: EGF/MAPK (core), WNT (both core and peripheral), TGF (core), and NHR.

In SignaLink each interaction is manually tagged with the PubMed ID of the original publication reporting experimental evidence for the interaction. This is necessary for selecting interactions with experimental evidence from the large set of 'common knowledge' (but often not sufficiently verified) interactions used in many reviews (Cusick et al., 2009). All interactions in SignaLink are directed. For most of the interacting protein pairs in SignaLink the unidirectional (only A→B) and for a few pairs it is bidirectional (both A→B and B→A). The interaction type descriptors ‘direct’ and ‘indirect’ indicate whether there is biochemical (direct) or other type of evidence (indirect), respectively. For details see the next section.

## *Manual curation process of SignaLink*

All pathways of the three species were compiled (*i.e.*, manually curated) separately. SignaLink contains no interologs (Cusick et al., 2009; Yu et al., 2004). For each pathway, we performed the following three main steps. First, we searched for pathway-specific review articles and databases. Next, we added to each pathway the signaling proteins listed as the members of that pathway in the review papers. Finally, we searched again for additional signaling interactions of the listed pathway proteins.

### Proteins

Data on proteins were collected as follows.

- ORFs and Ensembl protein IDs from species-specific databases: WormBase (version 191) (Rogers et al., 2008), FlyBase (version 2008.6) (Drysdale, 2008), and Ensembl (version 49) (Flicek et al., 2008).
- UniProt IDs from UniProt (version 87) (Boutet et al., 2007). If more than one UniProt ID was available for the same protein, then the ID(s) of the protein(s) with the longest amino acid sequence was (were) used.
- Protein name synonyms were listed from review papers, *e.g.*, Ref. (Kyriakis and Avruch, 2001), and the “synonym” field of the iHOP database (May 2008 version) (Fernandez et al., 2007).
- For protein ID conversions we used also PICR, the Protein Identifier Cross-Reference Service (Cote et al., 2007) and the Synergizer (Berriz and Roth, 2008).
- For each protein we listed its orthologs in the other two species with the help of the ortholog clusters of the InParanoid database (version 6.1) (Berglund et al., 2008). Consider an InParanoid cluster containing protein $P_A$ from species A and proteins $P_{B,1}$ and $P_{B,2}$ from species B. From the $P_B$ proteins in the cluster we included into SignaLink only those with an InParalog score equal to or above 0.3.
- When inserting a protein into SignaLink we assigned it to one pathway and – within that pathway – to one pathway region (*e.g.*, EGF/MAPK core, TGF non-core, or NHR core). Later further pathways and pathway regions were added for this protein, if necessary.
    - We marked a protein a ‘core’ component of a pathway, if it is essential for transmitting the signal of its pathway and has at least one of the pathway’s biochemical characteristics, *e.g.*, ‘Ser/Tyr-kinase activity’.
    - ‘Non-core’ (same as ‘peripheral’) indicates that the protein does not belong to the core of the pathway, but it modulates the pathway’s core proteins.
- We determined the pathway section(s) of each protein separately, *e.g.*, ligand or receptor. A maximum of two sections per protein were allowed.
    - The pathway position **ligand** indicates that the protein starts the signal of its pathway.
    - A **receptor** is the immediate receiver of this signal.
    - A **mediator** is a member of the pathway mediating the signal from the receptors towards transcription factors.

- A **co-factor** modulates the function of any other protein. Note that co-factors are often in the peripheral (non-core) region of their pathways.
  - **Transcription factor**: (i) transmits the signal received from its pathway towards another transcription factor (TF) protein, or (ii) forms a complex with other TF proteins, or (iii) binds to a specific promoter region on the DNA.
  - **Other**: non-signaling proteins with the roles of cellular motion, transport, and membrane anchoring. In the worm, the fly, and humans SignaLink contains 14, 10, and 13 proteins, respectively annotated with the pathway position "other". Out of these proteins 3, 1, and 5, respectively are co-factors at the same time too, and the remaining 11, 9, and 8 proteins have no additional pathway position, only "other". In summary, in SignaLink the pathway position "other" is used rarely, and it appears in combination with co-factor only.
  - **Unknown**: the position of this signaling protein in its pathway is not known.
- After compiling the lists of proteins and signaling interactions we performed two further steps:
  - We went through the list of proteins again and for each protein, P, we listed all those interacting proteins, R, for which SignaLink contains a P→R directed interaction. We added to the list of pathway memberships of P all pathways of each of these (R) proteins.
  - For each signaling protein in one species (A) we listed its orthologs from the other two species (B, C). For each of these orthologs we checked whether it has been found to interact with signaling proteins already listed in SignaLink (for species B or C). If yes, then we inserted the ortholog into SignaLink too.

## Interactions

We used the following rules for collecting data on interactions.

- We collected the signaling interactions of a protein from primary research articles listed in
  - review papers,
  - species-specific databases (FlyBase, WormBase) and UniProt,
  - iHOP, ChiliBot (Chen and Sharp, 2004), and PubMed search results.
- Details:
  - Each interaction inserted into SignaLink had to be directed.
  - We made an effort to find the earliest possible experimental article describing the interaction.
  - For all review papers and approximately 70% of all primary experimental articles, we examined not only the abstract, but also the main text.
  - In SignaLink the experimental *evidence* for a signaling interaction between the two involved proteins *is either direct or indirect*:
    - *Direct experimental evidence.* There is published biochemical evidence for the signaling interaction between the two proteins.
    - *Indirect experimental evidence.* There is no direct biochemical evidence for an interaction, however, published experimental results suggest that the interaction is very likely possible. Evidence types accepted here are (1) changes in mRNA/protein levels, enzyme activities, concentrations of the products of catalyzed reactions, and (2) docking domain structures.
  - The *effect of an interaction* can be activating or inhibitory. For interactions with indirect evidence, ++ and -- mean activating, while +- and -+ mean inhibitory interaction. A unidirectional interaction (A and B interact as either

A→B or B→A) has only one type of effect, but for the few bidirectional interactions (A→B and B→A are both present) more than one type of effect is possible between the two proteins.
  - Two signaling interactions between the same two proteins in opposing directions are listed separately in SignaLink.

## Pathways and proteins included into and excluded from SignaLink

Signal transduction – briefly summarized – is the transmission of extracellular signals towards the nucleus. Based on this definition some protein families, pathways, and cellular systems were fully or partially excluded from SignaLink.

Based on the summary of Gerhart (Gerhart, 1999) Pires da Silva and Sommer selected (Pires-daSilva and Sommer, 2003) seven signaling pathways that are continuously active during development, have an essential role after the developmental processes, and are conserved in metazoans. We selected the same seven pathways and separated the RTK pathway group to EGF/MAPK and Ins/IGF.

Several further signaling pathways used by Gerhart (for example, the integrin pathway or the apoptosis pathway), are active only occasionally (not continuously) and contribute mainly to adult physiology. These were excluded by Pires da Silva and Sommer. We excluded these pathways from SignaLink too. In addition to their late activity, the excluded pathways are often controlled by input signals that are not extracellular**.** Examples are the TOR and apoptosis pathways, some immune pathways (*e.g.*, the NF-kB pathway), and cell cycle-related pathways such as the p53 system.

For an additional protein to be included into SignaLink during the manual curation process, we required that (1) it has an experimentally verified role in one of the eight curated pathways or (2) it has a signaling interaction with an already listed SignaLink protein.

The following cellular systems perform tasks other than signal transduction, thus only those of their proteins are present in SignaLink that have a direct connection to signaling proteins:

- Clathrin- and non-clathrin-mediated endocytosis
- Nuclear import/export system
- Chromatin remodelling
- The ubiquitin system

Interactions between transcription factors (TFs) are mostly well-separated from signal transduction (ST) due to their different time scale, mechanisms, and cellular location (Papin et al., 2005). Thus, most TF-TF interactions have been omitted from SignaLink. Note that signal transduction pathways end on TFs, thus, many proteins are contained by both the ST system and the transcription regulation system. We stopped at the first TF of a pathway and did not insert its neighbors into SignaLink. If we found that several TFs directly interact with non-TF signaling proteins and these TFs interact themselves, then we inserted these TF-TF interactions into SignaLink too (*e.g.*, between STAT3 and SMAD3). However, the co-factors of the NHR pathway were excluded from SignaLink, because they are members of the TF network and not the ST system.

From the following protein families only those members have been included into SignaLink for which we could identify an experimentally verified role in at least one of the eight selected signaling pathways or an interaction with a SignaLink protein:

- Heterotrimeric G proteins,
- Non-receptor tyrosine kinases,
- Receptor tyrosine kinases (RTK).

From the 58 currently known RTK receptors SignaLink contains only 4 EGF and 4 IGF receptors (4 EGFR, 2 IGFR, IRR, and INSR) for the following reasons.

- From the RTK pathway group, EGF/MAPK and IGF/INS are two pathways with sufficiently large sizes for comparisons.
- Inserting all 58 RTK receptors and associated proteins (approximately 3-4 ligands and mediators per receptor) would merge the currently largest pathway, EGF/MAPK, with Ins/IGF, resulting in a single pathway significantly larger than all other pathways (Hh, JAK/STAT, NHR, Notch, TGF, and WNT) combined. Thus, inserting all RTK receptors and associated proteins would strongly reduce the usability of pathway comparisons.
- In experimental work EGF and IGF receptors are often discussed separately. They are considered to be the two most important and representative receptor groups of the RTK family. Within the RTK pathway system EGF and IGF receptors are two of the most frequently researched receptor groups and they have clearly different functional roles.

Note that despite their clearly different roles, the downstream effects of the two receptor groups (EGF and IGF) do have similarities: IGF receptors can also forward signals to the MAPK section of the EGF/MAPK pathway (Robinson et al., 2000).

## Curation process: Example

As an example we discuss here the human Notch pathway and one of its components, the human NOTCH1 protein.

- According to pathway-based reviews (Baron, 2003), in humans there are 4 Notch family member proteins: NOTCH1, NOTCH2, NOTCH3, and NOTCH4.
- Notch proteins have a specific role in transmitting signals (Weinmaster, 1997).
- The ligands and transcription factors between which Notch proteins transmit signals as well as several further proteins tuning the functions of Notch proteins are known (Bray, 2006).

### Protein data

- Most databases and review papers do not differentiate between splice variants. For the human NOTCH1 protein Ensembl (Flicek et al., 2008) contains two splice variants: ENSP00000277541 and ENSP00000360765. From these two the InParanoid database (Berglund et al., 2008) contains only the first, ENSP00000277541, therefore, we inserted into SignaLink only this splice variant. For proteins that have more than one splice variant, but none of them is present in the InParanoid database, we inserted into SignaLink the splice variant that has a primary UniProt accession (AC), as listed by Ensembl version 49.
- From Ensembl we included into SignaLink the UniProt accession(s) of the protein and from UniProt we used the following data fields: the protein’s description, the protein reference – if it contained interaction data, – and the protein’s cellular component. In addition, data from the protein's description and interaction fields were manually tested in the primary publications for further information.

- According to Ref. (Ilagan and Kopan, 2007), NOTCH1 is a core protein of its pathway and it functions as a receptor, mediator, or transcription factor. However, when in the Notch pathway, NOTCH1 functions either a receptor or a transcription factor, thus, we have included into SignaLink only these two pathway sections for NOTCH1.
- To make SignaLink as complete as possible, we searched for orthologs of the human NOTCH1 protein. Orthologs without known signaling interactions became predicted pathway proteins in SignaLink.
- From the InParanoid database we identified the *C. elegans* and *D. melanogaster* orthologs of the human NOTCH1 protein (ENSP00000277541). (In several cases we searched by both the UniProt and Ensembl protein IDs in InParanoid to find the protein.) Interestingly, the human NOTCH1 has two orthologs in worm (LIN-12 and GLP-1), but only one ortholog in fly (the protein N). We inserted all three orthologs into SignaLink. We listed the species-specific protein IDs and the UniProt ACs of the orthologs from WormBase and FlyBase.
- For ligands and transcription factors interacting with NOTCH1 we followed the same steps.

**Interactions**

- We listed articles describing signaling interactions between NOTCH1 and other proteins by browsing through the references of the above mentioned review papers and by using the search engines iHOP (Fernandez et al., 2007) and ChiliBot (Chen and Sharp, 2004). The search engine iHOP allows the user to search for all abstracts with interactions containing NOTCH1. With ChiliBot the interactions between two selected proteins can be directly searched for.
- Consider, for example, the interaction between NOTCH1 and TACE/ADAM17. This interaction is described in the experimental article (Brou et al., 2000). After reading the article we found that it describes (i) a putative cleavage site for TACE on NOTCH1 and (ii) a correlation between the in vitro enzymatic activity of TACE and the activity of NOTCH1. Thus, this article provides evidence for the activation of NOTCH1 by TACE.
- In addition, we searched directly for interactions between the orthologs of TACE and NOTCH1 in the other two species.

## *Comparing the approach of SignaLink with other approaches*

Signaling pathways are the building blocks of signaling networks, however, their definition is not yet standardised. Thus, currently used definitions of signaling pathways are based on a variety of functional, structural, and cell specific classifications (Bader et al., 2006). As a consequence of this, more than one definition is in use for the system-level examination of signaling pathways. For example, the definition of cross-talks has not yet been standardised.

Most of these approaches have a specific purpose, *e.g.***,** the characterisation of cell fate signals (Tseng and Dustin, 2002), the analysis of selected cell types (Grunwald and Klein, 2002) or selected pathways (Fuerer et al., 2008). Supplementary Table 2 lists several approaches to cross-talk analysis in the increasing order of the complexity of the examined system. The table starts with general cross-talk reviews of one selected pathway to intercellular cross-talk studies. While previous studies compared and discussed the cooperation of pathway pairs at most (Wang et al., 2009a; Nusse, 2003), SignaLink allows the user to perform both pathway-specific and system-level analyses for all eight pathways, with an emphasis on cross-talk.

Complexity ↓

| Focus of the analysis | Reference |
|---|---|
| **Cross-talks of**<br>in healthy cell types | (Guo and Wang, 2009)<br>(Hurlbut et al., 2007) |
| one pathway in stem cells | (Katoh and Katoh, 2007) |
| two pathways | (Borisov et al., 2009; Wang et al., 2009a) |
| selected pathways in one tissue type | (Natarajan et al., 2006) |
| many pathways in one tissue type | (Sternberg, 2005) |
| many pathways in stem cells | (Blank et al., 2008) |
| many pathways in one organ | (Yan et al., 2004) |
| **Intercellular cross-talk**<br>between two pathways in one organ | (Boswell et al., 2008) |

**Supplementary Table 2.** Approaches to signaling cross-talk analysis in metazoans.

Databases usually agree on the ‘cores’ (backbones) of signaling pathways, but peripheries are often captured very differently. To circumvent the ambiguity of pathway boundaries, Gerstein and colleagues have embedded pathway ‘cores’ into the large-scale network of all known PPIs (Lu et al., 2007). Lu *et. al.* also point out that the individual signaling databases of pathways are often assembled by several professionals applying different collection criteria and thus, it can be misleading to merge such pathway collections into a single database and perform cross-talk analyzes on them. The features setting SignaLink clearly apart from other pathway databases are shown in Supplementary Table 3**.**

## *Comparing the human signaling pathways of KEGG, Reactome, NetPath, and SignaLink*

Manual curation projects are (i) inherently error-prone (Cusick et al., 2009) and (ii) require specific goals that define their curation criteria. Comparisons between both the curation protocols (Rzhetsky et al., 2009) and the resulting databases are essential. In the current paper we publish a database and some of its possible applications. Therefore, not the curation process, but the database itself – SignaLink – should be compared to others.

The number of databases collecting signaling interactions is constantly growing (Bader et al., 2006). However, many of these databases are proprietary, list fewer than 200 molecules, or only selected types of pathway components (*e.g.*, protein kinases). Even among the few databases passing these criteria – free for academic use, more than 200 molecules, contains all pathway component types – there are currently no gold standards of signal transduction pathways compiled with similar purposes and methods as SignaLink. Thus, SignaLink as a whole cannot be validated at this moment.

Due to the absence of appropriate gold standards we compared the human signaling pathways of SignaLink with those from 3 widely used pathway databases: KEGG, Reactome, and NetPath. For reference we compared the human signaling pathways of these three databases to each other too. Except where explicitly stated, for the comparisons we mapped all protein IDs to UniProt primary accessions (*e.g.*, P42224). For the tables and text below we converted UniProt primary accessions to HGNC symbols (*e.g.*, we mapped P42224 to STAT1).

As discussed below, there are several important differences between these databases:

- Availability of a detailed description of the curation method
- Data sources (none, research articles, or only reviews)
- Entities connected by the interactions differ: pathways, proteins, ortholog groups, protein families, *etc.*
- Usage of protein isoform names
- Protein complexes allowed or not as a source of binary interactions
- Additional attributes of the interactions (*e.g.*, directed/undirected)
- Type of protein identifier used (UniProt AC, RefSeq ID, *etc.*)
- Number of cross-talks
- Database-specific (overrepresented) molecular functions

We believe that a comprehensive analysis of these differences is essential when deciding about which database should be used for a particular application. Below we compare the curation methods and contents (proteins and interactions) of the four databases. First through pairwise comparisons between the databases, *e.g.*, KEGG-SignaLink or Reactome-SignaLink, and then in a summary encompassing all four databases.

| | KEGG | BIOCARTA | pSTIING | REACTOME | Science Signaling – STKE | NCI Nature – Pathway Interaction Database | HPRD-NetPath | SignaLink |
|---|---|---|---|---|---|---|---|---|
| **Number of species** | Many but most networks are based on humans | Humans and mouse | Humans, rat, mouse, fly, worm, and yeast | Many | Many | Only humans | Only humans | 3 (*C. elegans*, *Drosophila*, Humans) |
| **Source of pathway components** | Subjective manual curation from the literature | Subjective manual curation from the literature | Subjective manual curation from the literature | Reaction based manual curation | Subjective manual curation from the literature | Subjective manual curation from the literature | Subjective manual curation from HPRD | Many pathway reviews, further manual curation |
| **Source of interactions** | Only reviews | Mostly reviews | PubMed ID listed for each interaction | PubMed ID listed for each interaction | PubMed ID listed for some interactions (not all) | PubMed ID listed for each interaction | PubMed ID and cited reaction details listed for each interaction | PubMed ID listed for each interaction |
| **Type of interactions** | Directed | Directed | Directed | Directed | Directed | Directed | Undirected | Directed |
| **Number of signaling pathways (only differences from SignaLink are shown)** | 10 (no NHR) | Many | 12 (no WNT, TGF, Notch, IGF, Hh, NHR ) | 10 but not for every species (no Hh, NHR, JAK/STAT) | 49 canonical, 32 specific (no NHR) | 82 (no NHR, JAK/STAT) | 20 (no WNT, IGF, NHR, JAK/STAT) | 8 (EGF/MAPK, WNT, TGF, Notch, IGF, Hh, NHR, JAK/STAT) |
| **Definition of pathways** | Not defined | Not defined | Immune pathways | Not defined; uniform curation rules | Not defined | Not defined | Cancer or immune pathways | Biochemically based, important in development; uniform curation rules |
| **Pathways with (i) uniform curation criteria and (ii) readily available for cross-talk analysis** | Not possible | Not possible | Possible | Possible, but no global pathway view or common platform | Not possible | Not possible | Not possible | Possible |
| **Availability (cost, data structure, website)** | Free, web based, www.genome.jp/kegg | Free, but only diagram image and protein list, www.biocarta.com | Free, web based, pstiing.licr.org | Free, SQL dump, www.reactome.org | Free after registration, Web based, SBML upon request, stke.sciencemag.org | Free, PID XML and BioPAX, pid.nci.nih.gov | Free, in many network formats www.netpath.org | Free, SQL dump, SBML signalink.org |

**Supplementary Table 3.** Comparison of SignaLink with other manually curated signaling databases.

## KEGG and SignaLink: A comparison of isoforms and the definition of interacting entities

### Listing signaling interactions from KEGG pathway files

The Kyoto Encyclopedia of Genes and Genomes (Ogata et al., 1999) contains a large number of pathways, among them signal transduction and endocrine pathways. Out of the 8 pathways curated in SignaLink 7 are listed in KEGG: MAPK, WNT, Notch, Hedgehog, TGF-beta, Jak-STAT, and Insulin.

The pathways of KEGG contain seven types of entries: compound, group, map, ortholog, gene, enzyme, and other. From these we used only orthologs and their interactions. In KEGG each 'ortholog' entry is a group of proteins, which may contain more than one human protein. In KEGG an interaction connects two entries, *e.g.*, two ortholog groups, and – to the best of our knowledge – an interaction between two ortholog groups does not contain information about which proteins from the two ortholog groups are the actual interactors.

For each of the above mentioned seven pathways we first downloaded the corresponding PSI-MI formatted xml pathway file from ftp://ftp.genome.jp/pub/kegg/pathway/organisms/hsa in April 2009. These seven KEGG pathways contain four types of entries (usage in brackets): ortholog (373), map (40), group (15), and compound (11). We discarded 'map' (a linked pathway map, which denotes an entire pathway) and 'compound' (a chemical compound including, for example, glycan) entries, and retained 'ortholog' and 'group' (a complex of gene products, mostly a protein complex) entries.

In the seven downloaded xml formatted pathway files the name of an 'ortholog' type entry contained one or more 'KEGG ortholog set (ko)'. For example, in the file 04330_Notch.xml the type of entry number 3 is 'ortholog' and its name is a space-separated list containing three KEGG ortholog sets: "ko:K04505 ko:K04522 ko:K06060". Each of these three KEGG ortholog sets contains one human protein. We used the file hsa_ko.list to map these KEGG ortholog set identifiers to KEGG's *Homo sapiens* protein numbers. In the example we mapped ko:K04505 to hsa:5663, ko:04522 to hsa:5664, and we could not map ko:06060 (this ID was discarded). Next, we mapped KEGG's *Homo sapiens* protein numbers to UniProt accessions with the file hsa_uniprot.list, in the example for hsa:5663 two UniProt accessions were listed (B2R6D3 and P49768), we mapped hsa:5663 to P49768 and similarly, we mapped hsa:5664 to the second of the two listed UniProt accessions (B1AP21, P49810). Note that in both cases, the first UniProt AC is from TrEMBL, while the second is from Swiss-Prot. By checking a few examples in the file hsa_uniprot.list we concluded that if for a KEGG *Homo sapiens* protein number more than one UniProt accession (AC) is listed and at least one of these ACs is from Swiss-Prot, then the last AC is from Swiss-Prot. Thus, to use a more reliable protein AC – one from Swiss-Prot, if possible – we used always the last AC.

The 'group' entries in the downloaded xml files of KEGG pathways contain components which are themselves entries from the same xml file. We replaced each 'group' entry with the list of its 'ortholog' components and we replaced each of these ortholog components with the list of the UniProt accessions found as explained above. In summary each pathway entry that we kept for the analysis contained one or more protein name (in UniProt AC format).

KEGG pathways list directed interactions between entries. Each of the entries we used contained one or more human protein. We used only those interactions where the two connected entries were both among the ortholog or group entries selected above. For the seven pathways this selected list of directed interactions gave 285 nodes (protein groups) and 210 directed links between these groups. The total number of proteins in these groups was 429.

## Constructing the network of signaling interactions from KEGG pathways

In the previous subsection we explained how we listed the directed signaling interactions connecting groups of human proteins from the KEGG pathways MAPK (we compare it to the EGF/MAPK pathway of SignaLink), WNT, Notch, Hedgehog, TGF-beta, JAK/STAT, and Insulin. We used two alternatives to construct a network of directed signaling interactions from these data. We used the ‘matrix’ and ‘spoke’ methods, which are frequently applied for constructing binary protein-protein interaction lists from protein complex data (Orchard et al., 2007).

We called the two groups of proteins connected in KEGG by a directed signaling interaction a *source group* and a *target group*. For the first KEGG signaling network (‘KEGG matrix’) we connected with a directed signaling interaction each protein of the source group with each protein of the target group. For the second KEGG signaling network (‘KEGG spoke’) we selected one protein from the source group and one protein from the target group and added a directed signaling interaction only between these two proteins. Both protein groups contained UniProt primary accessions (allowing both Swiss-Prot and TrEMBL proteins). We picked the selected protein from a group by (i) putting accessions into alphabetical order and then (ii) picking either the first Swiss-Prot accession from the list or – if only accessions from TrEMBL were listed – the last accession.

## Comparing the human signaling pathways of KEGG and SignaLink

To compare KEGG with SignaLink, we considered the ‘KEGG matrix’ and ‘KEGG spoke’ networks (see the previous section) constructed from the KEGG pathways MAPK, WNT, Notch, Hedgehog, TGF-beta, JAK/STAT, and Insulin. Note in Supplementary Table 4 that the number of ‘SignaLink only’ proteins is well above the number of KEGG spoke ‘KEGG only’ proteins. Only in the ‘KEGG matrix’ network – which very likely contains many false positive links – is the number of signaling interactions comparable to that of SignaLink.

For the comparisons in Supplementary Table 4 we mapped first each protein name (both in KEGG and SignaLink) to a UniProt primary accession (AC). The comparisons labelled “all isoforms names separately” were performed on these lists. Next we mapped each UniProt primary AC to the HGNC gene name listed in UniProt or – if there was no HGNC name – the truncated ID (the ID minus the “_HUMAN” tag). Finally, in these name lists we removed from the name of each protein any of the following patterns, if they followed a number: (i) a single letter (for example, WNT10A was replaced with WNT10), (ii) a letter and a number (e.g., CSNK2A1 was replaced with CSNK2), and (iii) a letter, a number and a letter (e.g., CSNK1A1L was replaced with CSNK1). In addition to these rules MAPK protein names (*i.e.*, names starting with “MAPK”) were not changed (protein names starting with ”MAP” followed by zero or more numbers, a ”K” and then optional further characters were considered to be MAPK protein names). We performed the comparisons labeled “merged isoform names” (M) on these final name lists.

| | **Pathway** | **Proteins and interactions** M: isoform names merged, S: all isoforms listed separately | | | |
|---|---|---|---|---|---|
| | | **only in KEGG** | | **intersection:** both in KEGG ‘matrix’ and SignaLink | **only in SignaLink** |
| | | KEGG ‘matrix’ | KEGG ‘spoke’ | | |
| | | M (S) | M (S) | M (S) | M (S) |
| **Proteins** | EGF/MAPK | 85 (101) | 57 (59) | 63 (63) | 98 (99) |
| | Hh | 16 (20) | 4 (5) | 14 (13) | 25 (26) |
| | IGF | 65 (65) | 51 (51) | 9 (9) | 26 (28) |
| | JAK/STAT | 24 (24) | 14 (14) | 54 (55) | 55 (66) |
| | Notch | 10 (10) | 8 (8) | 14 (14) | 25 (25) |
| | TGF | 26 (26) | 19 (19) | 19 (19) | 64 (67) |
| | WNT | 52 (62) | 40 (43) | 32 (26) | 41 (49) |
| | Total all proteins | 196 (224) | 137 (145) | 210 (205) | 252 (278) |
| | multi-pathway proteins | 27 (34) | 19 (19) | 8 (7) | 36 (36) |
| **Inter-Actions** | Total | 1332 (1411) | 301 (307) | 199 (191) | 775 (799) |
| | Cross-talks (weighted number) | 214.1 (199.8) | 74.5 (71.5) | 11.3 (10.3) | 216.8 (214.9) |

**Supplementary Table 4.** Protein and interaction numbers in the seven signaling pathways curated by both KEGG and SignaLink. The label “only in KEGG” indicates proteins (or interactions) listed only in KEGG, “intersection” indicates proteins (or interactions) listed in the given pathway(s) of both databases, and “only in SignaLink” indicates names listed only in SignaLink. In KEGG each directed signaling interaction connects two groups of proteins (a source group and a target group). In the KEGG ‘matrix’ network for each directed interaction we added a directed link between each protein of the source group and each protein of the target group. In the KEGG ‘spoke’ network only one protein from the source group was connected with one protein from the target group. See the section “Analysis of cross-talk in each species” in “Supplementary Results” above for the definition of weighted link numbers and a detailed description. Pathways in which the removal of isoform names significantly reduces the number of proteins are highlighted with gray background. See text for the discussion of this table**.**

Removing isoform names (*i.e.*, merging name variants) has a significant effect on the lists of proteins in several pathways. We found that KEGG contains a significant number of isoforms in both the MAPK and WNT pathways, and SignaLink contains some in the WNT pathway. In addition, both databases contain many – but not the same – isoforms in the JAK/STAT pathway. Observe also in the ‘M’ columns (where isoform names are merged) of Supplementary Table 4 that the KEGG ‘matrix’ data set contains significantly more proteins than SignaLink in the IGF pathway (74 vs. 35) and notably more proteins (84 vs. 73) in the WNT pathway, while all other pathways are more populous in SignaLink. In total SignaLink and the KEGG matrix data set have similar numbers of proteins and almost the same (weighted) cross-talk numbers. However, the KEGG matrix data set – as all networks prepared with the matrix method – is very likely to contain many false positive interactions. Thus, in our opinion SignaLink contains significantly more biologically meaningful signaling interactions, including cross-talks, than KEGG.

### Conclusions for the comparison between KEGG and SignaLink

Considering the seven signaling pathways curated in both databases, we found the following differences between KEGG and SignaLink:

- With the most permissive approach – KEGG 'matrix' with isoforms – the selected 7 pathways of KEGG have 20% fewer proteins than the same 7 pathways in SignaLink (429 vs. 483 proteins). However, with the same approach KEGG has 50% more (undirected) interactions (1 502 vs. 990). This large number of links probably includes many that were introduced only by the 'matrix' method, and do not exist in reality.
- Although in SignaLink the weighted number of cross-talks is higher, KEGG has significantly more (50%) interactions between fewer (20%) proteins. This is mainly due to the higher number of interactions within the pathways of KEGG.
- Several pathways contain large numbers of isoform names. After merging isoform names, the total number of proteins contained exclusively by either of the two databases decreases, while the total number of proteins contained by both databases increases.

## Reactome and SignaLink: Complexes and database-specific molecular functions

We downloaded the files uniprot_2_pathways.stid.txt and homo_sapiens.interactions.txt in April 2009 from the web directory http://www.reactome.org/download/current of Reactome (Joshi-Tope et al., 2005). There are five signaling pathways that have been curated both in Reactome and in SignaLink: EGF (EGRF in Reactome and EGF/MAPK in SignaLink), IGF/Ins, Notch, TGF, and WNT.

From the first of the two data files (uniprot_2_pathways.stid.txt) we extracted the protein lists of the EGFR pathway ("Signaling by EGFR", to be compared to the EGF/MAPK pathway of SignaLink), the IGF/Ins pathway ("Signaling by Insulin receptor"), the Notch pathway ("Signaling by Notch"), the TGF pathway ("Signaling by TGF beta" merged with "Signaling by BMP"), and the WNT pathway ("Signaling by Wnt"). Next we converted protein identifiers to UniProt primary accessions. From the second of the two data files (homo_sapiens.interactions.txt) we listed all interactions between the proteins of the five pathways. Reactome lists binary interactions classified into four interaction types: 'neighbouring_reaction' (528 of them connect proteins of the five selected pathways), 'reaction' (535), 'indirect_complex' (349), and 'direct_complex' (950).

SignaLink contains binary protein-protein interactions, but no binary associations between two proteins inferred from the membership of the two proteins in the same protein complex. Reactome contains both. In Reactome 'direct_complex' and 'indirect_complex' are the labels indicating binary associations between two proteins inferred from the membership of the two proteins in the same protein complex. For a mathematically and biologically meaningful comparison of Reactome and SignaLink only true binary protein-protein interactions should be used. In Supplementary Table 5 this comparison is marked 'N' (Binary interactions of Reactome inferred from membership in the same protein complex are excluded). As Reactome interaction data are often used in the literature without a distinction between Reactome data types, we have also performed the comparison by including 'complex' associations: in Supplementary Table 5 'C' indicates that in this comparison the Reactome data set includes binary protein-protein associations derived from the membership of the two proteins in the same protein complex.

| | Pathway | Proteins and interactions<br>N: Reactome without complexes<br>C: Reactome with complexes | | |
|---|---|---|---|---|
| | | **only in Reactome**<br>**N** (C) | **both in Reactome and SignaLink**<br>**N** (C) | **only in SignaLink** |
| **Proteins** | EGF/MAPK | 28 (28) | 16 (16) | 146 |
| | IGF | 32 (32) | 7 (7) | 30 |
| | Notch | 2 (2) | 14 (14) | 25 |
| | TGF | 10 (11) | 19 (19) | 67 |
| | WNT | 5 (53) | 4 (5) | 70 |
| | **Total**<br>all proteins | **56 (105)** | **64 (65)** | **283** |
| | multi-pathway proteins | 6 (15) | 0 (2) | 39 |
| **Inter-actions** | **Total**<br>all interactions | **622 (1454)** | **67 (73)** | **615** |
| | cross-talks (weighted sum) | 39.3 (39.3) | 0 (0) | 216.5 |

**Supplementary Table 5.** Statistics of the five signaling pathways curated in both Reactome and SignaLink. Observe that (i) protein lists significantly differ and (ii) in each pathway SignaLink contains a larger number of proteins. Note that the removal of 'complex' interactions from Reactome reduces from 58 to 9 the number of its WNT pathway proteins that have at least one interaction with another WNT pathway protein. On the other hand the removal of 'complex' interactions leaves the list of cross-talks in Reactome's selected five pathways unchanged. Thus, 'complex' interactions in Reactome connect pairs of proteins within pathways, mostly within the WNT pathway. Note also that the the number of proteins contained by both databases is low.

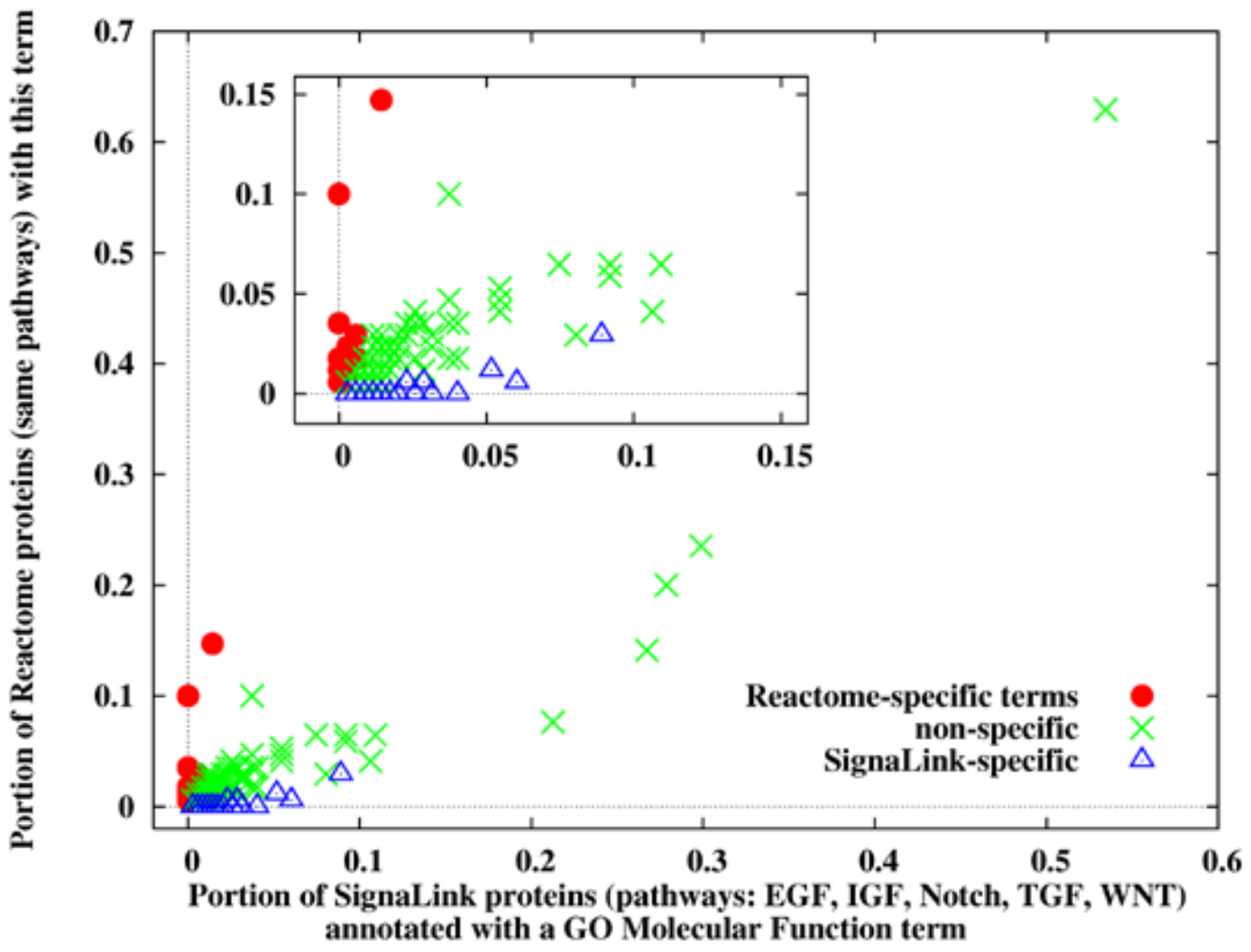

**Supplementary Figure 2.** The relative frequency of the usage of each GO Molecular Function term annotating the proteins of the five signaling pathways curated in both Reactome and SignaLink. The terms called "Reactome-specific" and "SignaLink-specific" are those terms that annotate at least three times as many proteins (compared to the total number of proteins) in one data set as in the other.

| GO Molecular Function term | Number of proteins annotated with this term | |
|---|---|---|
| | in Reactome | in SignaLink |
| **Reactome-specific terms** (at least 3 times more frequently used in the Reactome subset) | | |
| GO:0008233 peptidase activity | 25 | 5 |
| GO:0004298 threonine-type endopeptidase activity | 17 | - |
| GO:0016887 ATPase activity | 6 | - |
| GO:0043560 insulin receptor substrate binding | 5 | 2 |
| GO:0008601 protein phosphatase type 2A regulator activity | 5 | 2 |
| *etc.* | | |
| **SignaLink-specific terms** (at least 3 times more frequently used in the Reactome subset) | | |
| GO:0008083 growth factor activity | 5 | 31 |
| GO:0004713 protein tyrosine kinase activity | 2 | 18 |
| GO:0043565 sequence-specific DNA binding | 1 | 21 |
| GO:0004722 protein serine/threonine phosphatase activity | 1 | 8 |
| GO:0004708 MAP kinase kinase activity | 1 | 10 |
| GO:0004707 MAP kinase activity | 1 | 10 |
| *etc.* | | |
| **Non-specific terms** (most frequently used non-specific terms) | | |
| GO:0005515 protein binding | 107 | 186 |
| GO:0000166 nucleotide binding | 40 | 104 |
| GO:0005524 ATP binding | 34 | 97 |
| GO:0016740 transferase activity | 24 | 93 |
| GO:0005509 calcium ion binding | 17 | 13 |
| GO:0004674 protein serine/threonine kinase activity | 13 | 74 |
| *etc.* | | |

**Supplementary Table 6.** The usage of GO Molecular Function terms annotating the proteins of the five signaling pathways curated in both Reactome and SignaLink. The total number of proteins is 170 in the Reactome subset and 348 in the SignaLink subset.

In the five compared pathways both the protein and interaction lists of Reactome and SignaLink significantly differ (Supplementary Table 5). To quantify the possible reasons for this difference we first listed all Gene Ontology (GO) Molecular Function terms (Harris et al., 2004) that annotate at least one protein in Reactome or SignaLink. For each of these terms we counted the number of proteins it annotates in Reactome and the number of proteins it annotates in SignaLink. Supplementary Figure 2 and Supplementary Table 6 show these two numbers for each GO Molecular Function term and the two sets of terms typical for Reactome and SignaLink.

## Conclusions for the comparison between Reactome and SignaLink

In summary, we found the following main differences between the five signaling pathways curated in both Reactome and SignaLink:

- The selected five pathways of SignaLink contain more than twice as many proteins (348) as the same pathways in Reactome (170). After removing interactions labelled ‘direct_complex’ and ‘indirect_complex’ from Reactome, this difference becomes even larger: the five selected pathways of Reactome will contain only 56 such proteins that have at least one signaling interaction with another signaling protein in these five pathways.
- Without complexes Reactome has almost the same number of interactions in the five pathways as SignaLink (689 vs. 682). However, SignaLink has almost three times as many cross-talks (226 vs. 79). After adding Reactome's complex interactions ('direct_complex' and 'indirect_complex') the number of its within-pathway interactions increases more than twofold (to 1527), however, the number of cross-talks is unchanged. Thus, the 'complex' interactions of Reactome (*i.e.*, the binary signaling interactions inferred from the membership of two proteins in the same complex) are mostly within pathways.

- In Reactome the WNT pathway is dominated by interactions inferred from the membership of two proteins in the same complex.
- Several binding functions frequently occur in both databases. However when comparing the two databases to each other we found that proteolytic functions are more frequent in Reactome and kinases are more frequent in SignaLink.

## NetPath and SignaLink: Interactions, protein identifiers, cited papers, and curation approaches

NetPath was originally published as a part of the Human Protein Reference Database (Keshava Prasad et al., 2009) and was recently published in a separate database article (Kandasamy et al., 2010). It contains human proteins and interactions grouped into 20 pathways. In NetPath a protein is allowed to belong to more than one signaling pathway. There are 5 signaling pathways that are curated in both NetPath and SignaLink: EGFR1, Hedgehog, Notch, TGF beta Receptor, and WNT. The other 15 pathways in NetPath are specific for immune cells or cancers.

For each interaction NetPath provides a list of interactors. To convert these interactions to a list of interacting pairs, we used the 'matrix' model (Bader and Hogue, 2002): each pair of human proteins appearing together in at least one interaction was listed as an interacting pair. We discarded self-interactions, *i.e.*, interactions containing a protein interacting with itself. We note here that the number of human proteins in a NetPath interaction is typically two or three, thus, for the matrix model is a reasonably precise way to map the list of interactions (protein sets) to a list of interacting pairs.

For each of the five pathways we have

1. Downloaded pathway data in a single file from http://www.netpath.org/data/psimi_2,
2. Selected the human proteins (by the taxonomy ID: 9606) in each interaction,
3. Listed all human proteins of the pathway,
4. Listed pairs of interacting proteins with the 'matrix' model, and
5. Listed NetPath proteins with their RefSeq protein IDs, and mapped these later to UniProt primary accession (AC) identifiers.

For the comparisons shown in Supplementary Tables 7 and 8 we considered the proteins of the five selected pathways of NetPath and all interactions between these proteins. We did the same in SignaLink.

In Supplementary Table 8 we compare the lists of articles cited by the two databases for the five selected pathways. NetPath lists PubMed IDs (PMIDs) in each of its pathway files, *e.g.*, at http://www.netpath.org/data/psimi_2/NetPath_10_PSIMI.xml for the EGFR1 pathway. From SignaLink we listed for each pathway (or the five pathways together) the PMIDs of those articles that provided experimental evidence for the interactions connecting two proteins that are both contained by the selected pathway (or pathway group).

| Pathway | Number of proteins or interactions listed for this pathway | | |
|---|---|---|---|
| | only in NetPath | both in NetPath and SignaLink | only in SignaLink |
| **EGF/MAPK** | | | |
| **All proteins** | **115** | **19** | **143** |
| Listed in UniProt/Swiss-Prot | 72 | 19 | 143 |
| Listed in UniProt/TrEMBL | 42 | 0 | 0 |
| Not found in UniProt | 1 | 0 | 0 |
| **All interactions (undirected)** | **141** | **0** | **351** |
| **Hh** | | | |
| **All proteins** | **5** | **12** | **27** |
| Listed in UniProt/Swiss-Prot | 2 | 12 | 27 |
| Listed in UniProt/TrEMBL | 2 | 0 | 0 |
| Not found in UniProt | 1 | 0 | 0 |
| **All interactions (undirected)** | **10** | **4** | **53** |
| **Notch** | | | |
| **All proteins** | **42** | **11** | **28** |
| Listed in UniProt/Swiss-Prot | 25 | 11 | 27 |
| Listed in UniProt/TrEMBL | 16 | 0 | 0 |
| Not found in UniProt | 1 | 0 | 1 |
| **All interactions (undirected)** | **38** | **2** | **58** |
| **TGF** | | | |
| **All proteins** | **103** | **17** | **69** |
| Listed in UniProt/Swiss-Prot | 63 | 17 | 68 |
| Listed in UniProt/TrEMBL | 36 | 0 | 1 |
| Not found in UniProt | 4 | 0 | 0 |
| **All interactions (undirected)** | **186** | **0** | **188** |
| **WNT** | | | |
| **All proteins** | **61** | **20** | **55** |
| Listed in UniProt/Swiss-Prot | 36 | 20 | 55 |
| Listed in UniProt/TrEMBL | 20 | 0 | 0 |
| Not found in UniProt | 5 | 0 | 0 |
| **All interactions (undirected)** | **84** | **4** | **94** |
| **Five pathways together** | | | |
| **All proteins** | **281** | **81** | **274** |
| Listed in UniProt/Swiss-Prot | 171 | 81 | 273 |
| Listed in UniProt/TrEMBL | 100 | 0 | 1 |
| Not found in UniProt | 10 | 0 | 0 |
| Multi-pathway proteins | 15 | 2 | 39 |
| **All interactions (undirected)** | **446** | **11** | **690** |
| **Cross-talks (weighted sum)** | **58.3** | **1.5** | **169.6** |

**Supplementary Table 7.** Statistics of the five signaling pathways curated in both NetPath and SignaLink. For the comparison all protein names – RefSeq IDs from NetPath and UniProt accessions (AC) from SignaLink – have been converted to UniProt primary accessions. For the five signaling pathways NetPath and SignaLink list roughly the same number of proteins: 362 and 355, respectively. Altogether the two databases list 636 proteins for the five pathways, but only 81 protein names are listed in both cases. This large difference is in part due to the significant number of proteins in NetPath that have either not yet been verified fully with experiments and are listed in UniProt/TrEMBL (100 such proteins) or are not listed in UniProt at all (10 proteins). The number of cross-talks is much larger in SignaLink (weighted number: 171.1 vs. 59.8) and only a few of these cross-talks are present in both databases (weighted number: 1.5). Observe that in SignaLink the amount of proteins that belong to more than one pathway is more than twice as many as in NetPath. Note also that the low number of proteins contained by both NetPath and SignaLink makes the number of interactions contained by both databases also very small.

| Pathway | Number of papers cited for the interactions of this pathway | | |
|---|---|---|---|
| | only in NetPath | both in NetPath and SignaLink | only in SignaLink |
| **EGF/MAPK** | 132 | 3 | 206 |
| **Hh** | 9 | 5 | 23 |
| **Notch** | 47 | 10 | 24 |
| **TGF** | 95 | 18 | 74 |
| **WNT** | 53 | 17 | 61 |
| **All 5 pathways** all papers | 327 | 61 | 337 |

**Supplementary Table 8.** The number of articles cited for the interactions of the five signaling pathways (EGF/MAPK, Hedgehog, Notch, TGF, and WNT) that are curated in both NetPath and SignaLink.

## Conclusions for the comparison between NetPath and SignaLink

We conclude that the major differences between the five signaling pathways curated in both NetPath and SignaLink are the following:

- SignaLink contains more than two times as many multi-pathway proteins and nearly three times as many cross-talks as NetPath.
- NetPath – compared to SignaLink – contains a much higher number of proteins listed only in UniProt/TrEMBL, but not in UniProt/Swiss-Prot. Note that proteins listed in UniProt/TrEMBL have not yet been fully experimentally verified as known proteins.
- The two databases (NetPath and SignaLink) were compiled with very different concepts in mind. We assume that NetPath was compiled using protein properties listed in UniProt (both Swiss-Prot and TrEMBL) and interactions in HPRD. As a contrast, SignaLink is review-based and its concept is to identify the functions of known proteins within their signaling pathways. This conceptual difference between NetPath and SignaLink may partly explain why the number of papers used for the compilation of both NetPath and SignaLink is low (see Supplementary Figure 8).
- SignaLink contains directed binary signaling interactions, while NetPath lists groups of interacting signaling proteins.
- SignaLink is available for three metazoan species, while NetPath is available only for humans.

## Pairwise comparisons between all four databases: Reactome, NetPath, KEGG, and SignaLink

As shown in the previous sections the four compared databases list significantly different numbers of proteins for several signaling pathways. The following signaling pathways have been curated in all four databases: EGF/MAPK (MAPK, EGFR1), Notch, TGF, and WNT. In Supplementary Figure 3 we compare all four databases on one diagram using these four pathways.

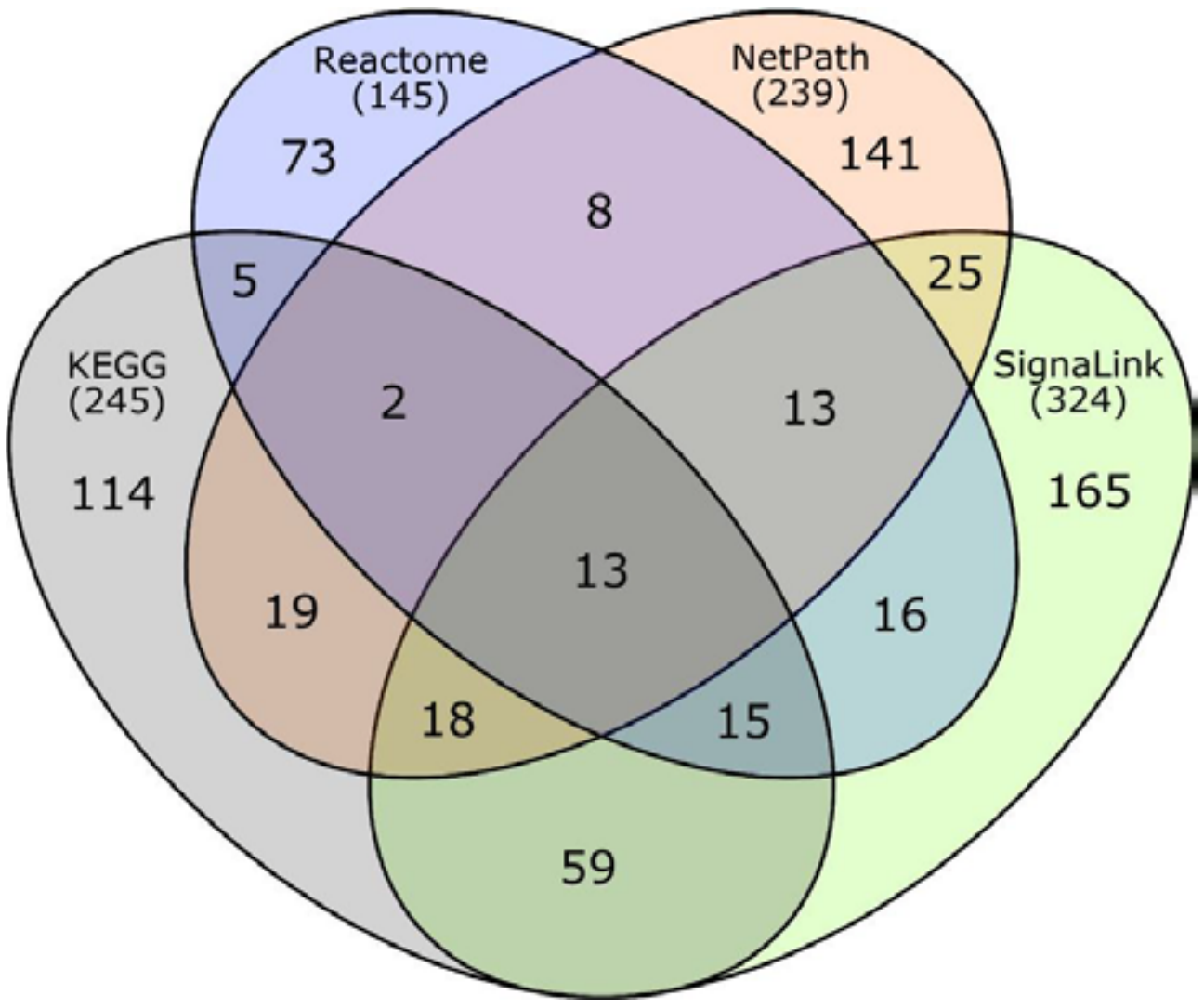

**Supplementary Figure 3.** Venn diagram comparing the four signaling pathways (EGF/MAPK, Notch, TGF, and WNT) curated in all four databases: Reactome, NetPath, KEGG, and SignaLink. Each section of the diagram shows the number of proteins in that set. The total number of proteins in the four pathways of each database is shown below the name of the corresponding database. Only proteins listed in UniProt/Swiss-Prot are included (proteins listed in UniProt/TrEMBL have been excluded from this comparison). Diagram layout from http://en.wikipedia.org/wiki/Venn_diagram.

## Comparing the lists of publications referenced in signaling pathway databases

| Database name | Number of all pathways (signaling and other) | Number of human pathways that have an equivalent in SignaLink | Number of equivalent human pathways in SignaLink |
|---|---|---|---|
| BioCarta | 351* | 34 | 8 |
| KEGG | 270** | 7 | 7 |
| NCI | 107* | 20 | 5 |
| NetPath | 20* | 5 | 5 |
| pSTIING | 38* | 10 | 3 |
| Reactome | 997* | 5 | 5 |
| SignaLink | 8** | 8 | 8 |
| STKE | 84** | 35 | 8 |

**Supplementary Table 9a.** Number of human pathways in cellular signaling databases. *Only human pathways. **A pathway available in more than one organism is counted as one pathway.

| Database name | Total number of cited papers (signaling and other pathways) | Number of cited papers for the signaling pathways available in SignaLink | |
|---|---|---|---|
| | | Number of cited papers | Average number per pathway |
| BioCarta* | 1174 | 86 | 2 |
| KEGG | 2388 | 73 | 10.4 |
| NCI* | 1287 | 200 | 10 |
| NetPath* | 1109 | 351 | 70.2 |
| pSTIING[†,*] | n.a. | n.a. | n.a. |
| Reactome | 4491 | 166 | 33.2 |
| SignaLink[‡] | 941 | 941 | 42.8 |
| STKE[†] | n.a. | n.a. | n.a. |

**Supplementary Table 9b.** Statistics of the publications cited in cellular signaling databases. Included are those publications that have been used for manual curation in the given pathway database and are available on its website either as a downloadable file or for viewing with a browser. Note that while Supplementary Table 8 counts the publications cited for the interactions of a given pathway in SignaLink, in this table the numbers in the last column indicate the publications cited for pathway membership in SignaLink.*Only human pathways. [†]No online source of the list of publications used for curation could be identified (n.a.: not available). [‡]To curate pathway membership and interactions SignaLink uses a total of 941 review and research articles while for curating pathway membership only review papers have been used.

In Supplementary Table 9b we compare the numbers of publications cited by BioCarta, KEGG, NCI (NCI-Nature Pathway Interaction Database), NetPath, pSTIING, Reactome, and STKE (Signal Transduction Knowledge Environment, Database of Cell Signaling) with the list of publications cited by SignaLink. We found that in SignaLink the number of PubMed citations used per pathway is above the average of the listed databases.

## Listing cited papers

For **BioCarta** we started from the URL http://www.biocarta.com/genes/allPathways.asp and downloaded the papers cited under "References" on the page of each pathway. From **NetPath** we downloaded and unzipped the file http://www.netpath.org/download/zipped/PSI-MI.zip and then listed pathways from this file. For **KEGG** we listed cited papers from the file ftp://ftp.genome.jp/pub/kegg/pathway/pathway. For the **NCI** database we started from the file http://pid.nci.nih.gov/browse_pathways.shtml and downloaded the references of all NCI-curated pathways (subnetworks excluded). For **pSTIING** we started from the URL http://pstiing.licr.org/search/a_start_pathway.jsp, but could not locate the PubMed identifiers of the articles used for manual curation. For **Reactome**, we first downloaded the MySQL dump of the full database from http://www.reactome.org/download/index.html, but could not find the PubMed IDs in it. However, it was possible to download the list of PubMed IDs from the BioMart server at http://www.reactome.org/cgi-bin/mart. In the case of **Science STKE** (Signal Transduction Knowledge Environment) we started at the following URL: http://stke.sciencemag.org/cgi/collection, but could not find the list of PubMed IDs used for manual curation of the pathways.

## Final conclusions from the database comparisons

It is widely assumed that so far in each investigated organism only a small portion of all protein-protein interactions – this includes signaling interactions – have been identified and most interactions remain to be mapped (Venkatesan et al, 2009). To minimise the number of false positives entered into SignaLink we have strictly adhered to the rules described above in the section titled “Manual curation process of SignaLink” during the curation process. Since the number of missing interactions is large in each signaling database and SignaLink was optimised towards minimizing the number of false positive interactions, we think that the high number of pathway proteins and the high number of interactions is one of the advantages of SignaLink.

SignaLink was compiled based on reviews and primary research articles, therefore, the high numbers of proteins, cross-talks, and multi-pathway proteins indicate a high coverage (true positives). It is increasingly recognized that signaling pathways are not isolated entities, but form a single pathway system, thus, cross-talks and multi-pathway proteins are especially informative, and a high number of these entities (true positives) is also an advantage. As of yet only rough estimates can be made for the number of signaling interactions in the three investigated organisms, but the detailed description of the curation process of SignaLink provides strong support for its higher coverage.

We note that Cusick et. al. (Cusick et al, 2009) have pointed out several drawbacks of manually curating “low-throughput” (LTP) data compared to collecting data with high-throughput (HTP) methods. For the specific case of intracellular signaling we add the following comments to the results of Cusick et. al. In intracellular signaling (i) an exceptionally large portion of interactions is localised to, *e.g.*, membranes, and cannot be efficiently detected with current HTP techniques, and (ii) the manual curation process of SignaLink eliminates some of the common disadvantageous features of manually curated databases (*e.g.*, absence of proper documentation for the interactions) that have been highlighted by Cusick *et.al*. (2009). Most importantly, the manual curation process of SignaLink is based on a combination of experimental (research) and review articles. In the first set of sources (original research articles) experimental evidence is published, while in the second set of sources (review articles) specialists analyze available evidence about the relevance and quality of experimental data.

Based on the comparisons between signaling databases, we found that compared to Reactome, NetPath, and KEGG the advantages of SignaLink are the following:

- Precisely defined and documented curation protocol
- Highest numbers of proteins and interactions for the curated signaling pathways
- Highest numbers of cross-talks and multi-pathway proteins
- Largest overlap with the other three databases
- Number of publications per pathway higher than average
- Minimal usage of isoforms
- No binary interactions inferred from membership in the same protein complex
- Low number of proteins for which there is not yet enough experimental evidence (these are proteins listed in UniProt/TrEMBL, not in UniProt/Swiss-Prot).

## *Protein expression in healthy tissue types*

### Selecting the expression data set: eGenetics

Two of the currently available major, up-to-date sources of human tissue- (and organ-) specific protein expression data are the GNF Atlas (Su et al., 2004) and the eGenetics data set (Kelso et al., 2003). Both are integrated into Ensembl. The GNF Atlas was produced in a single series of experiments and it contains protein expression for 46 human tissues, organs, and cell lines. The eGenetics data set is a collection of small-scale experiments (7016 cDNA libraries represented in dbEST and 104 SAGE libraries) with protein expression data in a total of 50 different human tissue and organ types. In their top 19 tissue types the GNF Atlas and eGenetics contain protein expression for an average of 251.6 and 242.4 SignaLink proteins, respectively. In other words, the two expression data sets provide approximately the same coverage of SignaLink. We decided to use eGenetics because its data were more easily accessible.

The eGenetics data set lists for each protein the tissue/organ types where that protein is expressed. This data set contains a total of 524 proteins from SignaLink. Out of these 323 are expressed in at least one tissue type (in the eGenetics ontology “anatomical system” means all tissue types). In Supplementary Table 10 we list the tissue and organ types where – according to eGenetics – the highest numbers of SignaLink proteins are expressed. This is the full list of tissue/organ types used in eGenetics (in alphabetical order): adrenal gland, alimentary, artery, auditory apparatus, basal nuclei, brain, brain stem, breast, cardiovascular, central nervous system, cerebellum, cerebral cortex, cerebrum, colorectal, dermal, diencephalon, endocrine, endocrine pancreas, epithalamus, female genitals, ganglion, haematological, heart, intestine, joint, liver and biliary system, lung, lymph node, lymphoreticular, male genitals, meninges, mesoderm, midbrain, muscle, musculoskeletal, omentum, oral cavity, osseous labyrinth, penis, peripheral nervous system, pharynx, respiratory, retina, skin, small intestine, subthalamus, urinary, urogenital, uterus, visual apparatus.

| eGenetics ontology | Number of expressed SignaLink proteins | Percent of human SignaLink proteins with pathway information present in eGenetics |
|---|---:|---:|
| anatomical system (all tissue/organ types) | 323 | 61.8% |
| respiratory | 303 | 57.9% |
| male genitals | 294 | 56.2% |
| female genitals | 293 | 56.0% |
| central nervous system | 288 | 55.1% |
| Alimentary | 264 | 50.5% |
| **Colorectal** | **261** | **49.9%** |
| Endocrine | 249 | 47.6% |
| Urinary | 247 | 47.2% |
| **Dermal** | **246** | **47.0%** |
| visual apparatus | 238 | 45.5% |
| Musculoskeletal | 237 | 45.3% |
| **liver and biliary system** | **236** | **45.1%** |
| Lymphoreticular | 235 | 44.9% |
| **Cardiovascular** | **230** | **44.0%** |
| Uterus | 220 | 42.1% |
| Haematological | 212 | 40.5% |
| peripheral nervous system | 190 | 36.3% |
| **Muscle** | **188** | **36.0%** |
| Breast | 174 | 33.2% |

**Supplementary Table 10.** The number of expressed signaling proteins (from SignaLink) in each human tissue/organ type according to eGenetics. The five tissue types selected for the analyzes are marked with dark background color.

## Selecting the five tissue types

Our goal was to check whether human cross-talks are selectively active, *i.e.*, whether cross-talks connecting different pathway pairs are active in different numbers of tissue types. To obtain a statistically reliable statement, we decided to analyze several tissue types together. On the other hand, using expression from all tissue types together would blur the cross-talk differences we would like to detect. As an optimum we selected five tissue types. We listed from the full eGenetics expression data set only protein expression in these five tissues and merged this reduced expression data set with SignaLink. The criteria for selecting the five tissue types were the following: (i) eGenetics should contain expression data for a high number of SignaLink proteins in these five tissues, *i.e.*, in eGenetics these five tissues should be as specific for SignaLink proteins as possible; (ii) each selected tissue type should be biologically homogeneous, and (iii) they should be of different origin (ecto-, meso-, entodermal) and function (*e.g.,* skin, liver).

To satisfy (i) we first listed the tissue types where the highest numbers of SignaLink proteins were expressed. Observe in Supplementary Table 10 that the first few tissue types on the list are rather heterogeneous: each of them is distributed among several organs, *e.g.*, male and female genitals or the alimentary and endocrine systems. To satisfy conditions (ii) and (iii), we looked on this list for those five tissue types that were the most homogeneous and of sufficiently different origin. Thus, from the top of the list we selected cardiovascular, colorectal, and dermal tissues, the liver/biliary system, and muscle. Supplementary Table 11 shows for each curated signaling pathway the portion of its proteins expressed in the five selected tissue types.

| Pathway | Average expression of the pathway's proteins in the five selected tissue types |
|---|---|
| Notch | 48.6% |
| IGF | 45.9% |
| TGF | 44.2% |
| EGF/MAPK | 42.3% |
| NHR | 38.8% |
| Hh | 34.2% |
| WNT | 33.3% |
| JAK/STAT | 25.8% |

**Supplementary Table 11.** Average percentage of the proteins of each human signaling pathway expressed in the five selected tissues types. For a given pathway we first computed the percent of its proteins expressed in each of the five selected tissue types and then we averaged this value over the five tissues.

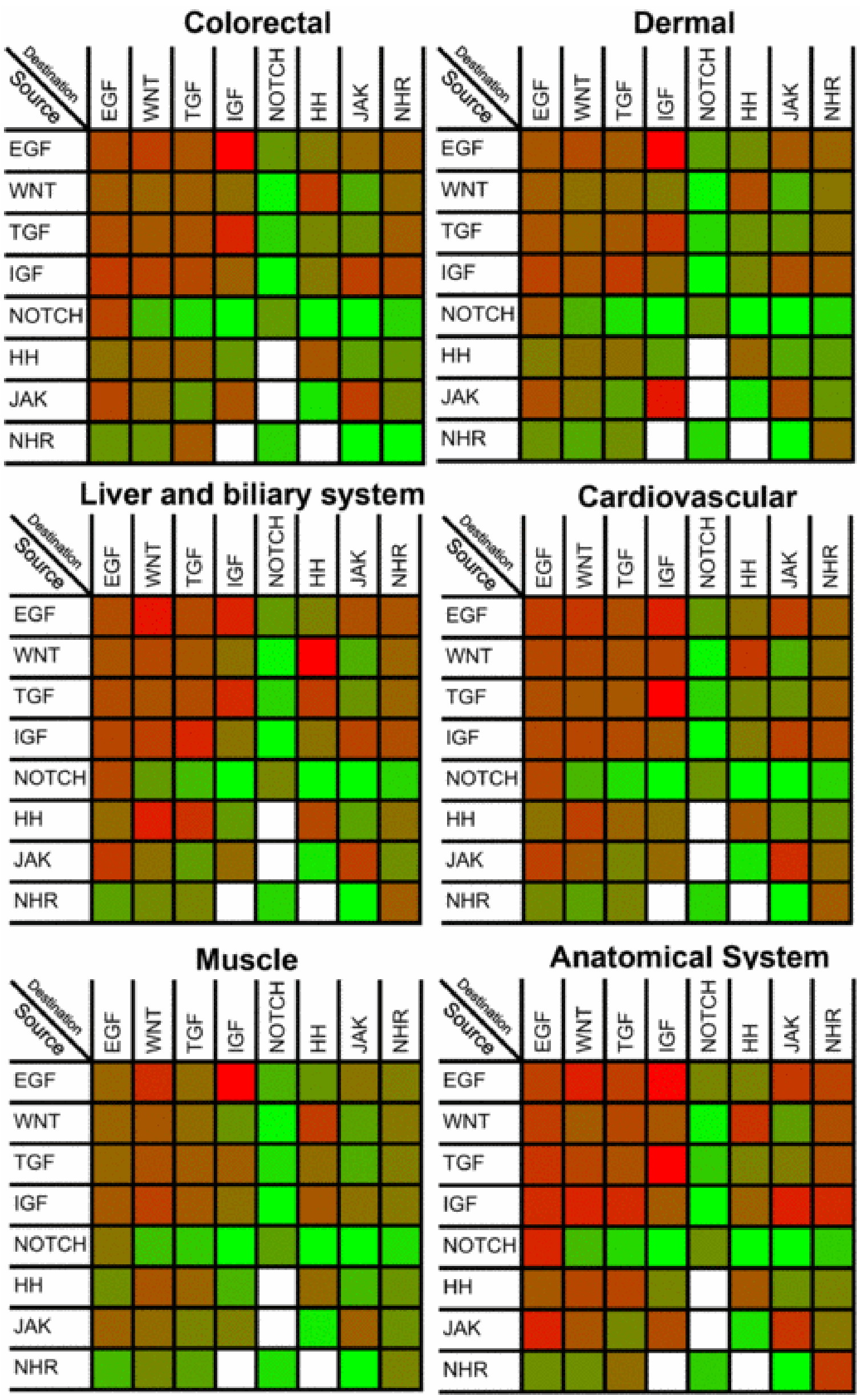

**Supplementary Figure 4.** Control for Figure 3a of the main text. See text for details.

## Controls

For Figure 3a of the main text we applied two types of controls shown in Supplementary Figure 4. First, we performed separately for each of the five selected tissue types the same analysis as in Fig. 3a. Second, we selected the category “anatomical system” (*i.e.*, all tissue types pooled). The main characteristics – discussed in the main text and in the supplementary material above – are preserved.

## *Statistical significance analyses of pathway member protein functions*

### Gene Ontology term enrichment tests in the curated signaling pathways

For a rigorous statistical analysis of the functions of proteins in the curated pathways and their overlaps we performed statistical enrichment tests of Gene Ontology Biological Process terms using the hypergeometric distribution and a (Bonferroni) multiple hypothesis correction. We performed this test in each of the signaling pathways and their overlaps with the 'GO Termfinder' toolbox, implemented by (Boyle et al., 2004). Note that GO TermFinder uses a list of GO terms not only as a plain list, but it makes of the Directed Acyclic Graph (DAG) of GO terms to quantify the significance of term enrichment in a group of proteins.

For each of the 22 signaling pathways available in SignaLink Supplementary Table 12 lists the 5 (for the EGF/MAPK pathway 6 or 7) most significant GO Biological Process terms. This analysis has pointed out that each pathway does have its own specific terms, many of which are explicitly are the GO terms named after the pathway, *e.g.*, MAPKKK cascade. In each of the 3 species the listed (most significant) GO terms of different pathways are different terms, in other words: the listed pathways are functionally distinct entities.

### *Caenorhabditis elegans*

| **Pathway in SignaLink** | **Most significant GO Biological Process terms of the proteins of the pathway** (with corrected P-values using a Bonferroni correction) | | |
|---|---|---|---|
| EGF/MAPK | GO:0040025 | vulval development | $7.3 * 10^{-31}$ |
| | GO:0007165 | signal transduction | $2.1 * 10^{-28}$ |
| | GO:0007154 | cell communication | $4.5 * 10^{-28}$ |
| | GO:0040028 | regulation of vulval development | $1.2 * 10^{-23}$ |
| | GO:0048580 | regulation of post-embryonic development | $1.2 * 10^{-23}$ |
| | … | | |
| | GO:0000165 | MAPKKK cascade | $6.7 * 10^{-16}$ |
| | … | | |
| IGF | GO:0040024 | dauer larval development | $4.7 * 10^{-16}$ |
| | GO:0042221 | response to chemical stimulus | $2.6 * 10^{-14}$ |
| | GO:0008286 | insulin receptor signaling pathway | $9.6 * 10^{-14}$ |
| | GO:0032868 | response to insulin stimulus | $9.6 * 10^{-14}$ |
| | GO:0032869 | cellular response to insulin stimulus | $9.6 * 10^{-14}$ |
| | … | | |
| NHR | GO:0006350 | transcription | 0 (for all five GO terms) The SignaLink NHR pathway contains all 246 *C. elegans* proteins in these 5 GO groups. |
| | GO:0006351 | transcription, DNA-dependent | |
| | GO:0006355 | regulation of transcription, DNA-dependent | |
| | GO:0009889 | regulation of biosynthetic process | |
| | GO:0010468 | regulation of gene expression | |
| | … | | |
| NOTCH | GO:0007219 | Notch signaling pathway | $1.1 * 10^{-19}$ |
| | GO:0008593 | regulation of Notch signaling pathway | $2.4 * 10^{-12}$ |
| | GO:0001708 | cell fate specification | $2.2 * 10^{-11}$ |
| | GO:0045165 | cell fate commitment | $1.6 * 10^{-10}$ |
| | GO:0045747 | positive regulation of Notch signaling pathway | $2.0 * 10^{-8}$ |
| | … | | |
| TGF | GO:0007178 | transmembrane receptor protein serine/threonine kinase signaling pathway | $6.4 * 10^{-15}$ |
| | GO:0007179 | transforming growth factor beta receptor signaling pathway | $3.3 * 10^{-12}$ |
| | GO:0007167 | enzyme linked receptor protein signaling pathway | $7.6 * 10^{-12}$ |
| | GO:0040024 | dauer larval development | $2.0 * 10^{-11}$ |
| | GO:0019219 | regulation of nucleobase, nucleoside, nucleotide and nucleic acid metabolic process | $1.9 * 10^{-10}$ |
| | … | | |
| WNT | GO:0045165 | cell fate commitment | $3.2 * 10^{-37}$ |
| | GO:0030154 | cell differentiation | $1.2 * 10^{-31}$ |
| | GO:0048869 | cellular developmental process | $7.7 * 10^{-31}$ |
| | GO:0001708 | cell fate specification | $2.5 * 10^{-28}$ |
| | GO:0050794 | regulation of cellular process | $5.8 * 10^{-25}$ |
| | … | | |

**Supplementary Table 12a.** The statistically most significant GO Biological Process terms for each signaling pathway of *C. elegans* curated in SignaLink. See the text above for details.

## *Drosophila melanogaster*

| **Pathway in SignaLink** | **Most significant GO Biological Process terms of the proteins of the pathway** (with corrected P-values using a Bonferroni correction) | | |
|---|---|---|---|
| EGF/MAPK | GO:0007165 | signal transduction | $3.9 * 10^{-39}$ |
| | GO:0007154 | cell communication | $1.1 * 10^{-35}$ |
| | GO:0050794 | regulation of cellular process | $8.7 * 10^{-34}$ |
| | GO:0050789 | regulation of biological process | $9.2 * 10^{-33}$ |
| | GO:0007167 | enzyme linked receptor protein signaling pathway | $5.1 * 10^{-32}$ |
| | … | | |
| | GO:0000165 | MAPKKK cascade | $4.9 * 10^{-24}$ |
| | GO:0007173 | epidermal growth factor receptor signaling pathway | $6.5 * 10^{-24}$ |
| | … | | |
| HH | GO:0007224 | smoothened signaling pathway | $1.2 * 10^{-47}$ |
| | GO:0008589 | regulation of smoothened signaling pathway | $1.2 * 10^{-29}$ |
| | GO:0007166 | cell surface receptor linked signal transduction | $5.6 * 10^{-27}$ |
| | GO:0007165 | signal transduction | $9.0 * 10^{-23}$ |
| | GO:0007154 | cell communication | $1.1 * 10^{-20}$ |
| | … | | |
| IGF | GO:0008286 | insulin receptor signaling pathway | $4.5 * 10^{-36}$ |
| | GO:0032868 | response to insulin stimulus | $4.5 * 10^{-36}$ |
| | GO:0032869 | cellular response to insulin stimulus | $4.5 * 10^{-36}$ |
| | GO:0043434 | response to peptide hormone stimulus | $4.5 * 10^{-36}$ |
| | GO:0032870 | cellular response to hormone stimulus | $8.3 * 10^{-34}$ |
| | … | | |
| JAK-STAT | GO:0007259 | JAK-STAT cascade | $8.1 * 10^{-26}$ |
| | GO:0007243 | protein kinase cascade | $2.2 * 10^{-18}$ |
| | GO:0046425 | regulation of JAK-STAT cascade | $1.2 * 10^{-16}$ |
| | GO:0007242 | intracellular signaling cascade | $2.4 * 10^{-12}$ |
| | GO:0010627 | regulation of protein kinase cascade | $3.3 * 10^{-9}$ |
| | … | | |
| NHR | GO:0006355 | regulation of transcription, DNA-dependent | $2.9 * 10^{-20}$ |
| | GO:0051252 | regulation of RNA metabolic process | $1.5 * 10^{-19}$ |
| | GO:0006351 | transcription, DNA-dependent | $2.2 * 10^{-19}$ |
| | GO:0032774 | RNA biosynthetic process | $2.3 * 10^{-19}$ |
| | GO:0045449 | regulation of transcription | $6.9 * 10^{-19}$ |
| | … | | |
| NOTCH | GO:0007219 | Notch signaling pathway | $2.5 * 10^{-43}$ |
| | GO:0007166 | cell surface receptor linked signal transduction | $1.8 * 10^{-26}$ |
| | GO:0007154 | cell communication | $3.6 * 10^{-25}$ |
| | GO:0007165 | signal transduction | $2.2 * 10^{-24}$ |
| | GO:0008593 | regulation of Notch signaling pathway | $2.2 * 10^{-19}$ |
| | … | | |
| TGF | GO:0007179 | transforming growth factor beta receptor signaling pathway | $1.5 * 10^{-30}$ |
| | GO:0007178 | transmembrane receptor protein serine/threonine kinase signaling pathway | $2.6 * 10^{-28}$ |
| | GO:0007167 | enzyme linked receptor protein signaling pathway | $1.1 * 10^{-19}$ |
| | GO:0007165 | signal transduction | $2.4 * 10^{-18}$ |
| | GO:0007166 | cell surface receptor linked signal transduction | $2.8 * 10^{-17}$ |
| | … | | |
| WNT | GO:0016055 | Wnt receptor signaling pathway | $8.8 * 10^{-54}$ |
| | GO:0007166 | cell surface receptor linked signal transduction | $1.7 * 10^{-34}$ |
| | GO:0007165 | signal transduction | $4.6 * 10^{-27}$ |
| | GO:0030111 | regulation of Wnt receptor signaling pathway | $5.6 * 10^{-25}$ |
| | GO:0007154 | cell communication | $1.2 * 10^{-24}$ |
| | … | | |

**Supplementary Table 12b.** The statistically most significant GO Biological Process terms for each signaling pathway of *D. melanogaster* curated in SignaLink. See the text above for details.

## *Homo sapiens*

| **Pathway in SignaLink** | **Most significant GO Biological Process terms of the proteins of the pathway** (with corrected P-values using a Bonferroni correction) | | |
|---|---|---|---|
| EGF/MAPK | GO:0007243 | protein kinase cascade | $5.8 * 10^{-88}$ |
| | GO:0006793 | phosphorus metabolic process | $1.8 * 10^{-80}$ |
| | GO:0006796 | phosphate metabolic process | $1.8 * 10^{-87}$ |
| | GO:0007242 | intracellular signaling cascade | $1.0 * 10^{-85}$ |
| | GO:0016310 | phosphorylation | $1.8 * 10^{-83}$ |
| | … | | |
| | GO:0000165 | MAPKKK cascade | $4.8 * 10^{-76}$ |
| | … | | |
| HH | GO:0007224 | smoothened signaling pathway | $1.4 * 10^{-15}$ |
| | GO:0009953 | dorsal/ventral pattern formation | $2.1 * 10^{-13}$ |
| | GO:0007275 | multicellular organismal development | $4.9 * 10^{-13}$ |
| | GO:0032502 | developmental process | $1.1 * 10^{-11}$ |
| | GO:0048856 | anatomical structure development | $5.5 * 10^{-11}$ |
| | … | | |
| IGF | GO:0032870 | cellular response to hormone stimulus | $1.7 * 10^{-26}$ |
| | GO:0008286 | insulin receptor signaling pathway | $3.0 * 10^{-25}$ |
| | GO:0032869 | cellular response to insulin stimulus | $1.2 * 10^{-24}$ |
| | GO:0032868 | response to insulin stimulus | $1.6 * 10^{-24}$ |
| | GO:0043434 | response to peptide hormone stimulus | $8.2 * 10^{-24}$ |
| | … | | |
| JAK-STAT | GO:0007259 | JAK-STAT cascade | $1.3 * 10^{-60}$ |
| | GO:0018108 | peptidyl-tyrosine phosphorylation | $5.7 * 10^{-48}$ |
| | GO:0018212 | peptidyl-tyrosine modification | $1.2 * 10^{-47}$ |
| | GO:0007243 | protein kinase cascade | $1.2 * 10^{-44}$ |
| | GO:0007260 | tyrosine phosphorylation of STAT protein | $2.0 * 10^{-41}$ |
| | … | | |
| NHR | GO:0006355 | regulation of transcription, DNA-dependent | $5.3 * 10^{-49}$ |
| | GO:0051252 | regulation of RNA metabolic process | $1.7 * 10^{-48}$ |
| | GO:0006351 | transcription, DNA-dependent | $1.6 * 10^{-47}$ |
| | GO:0032774 | RNA biosynthetic process | $1.7 * 10^{-47}$ |
| | GO:0016070 | RNA metabolic process | $2.1 * 10^{-41}$ |
| | … | | |
| NOTCH | GO:0007219 | Notch signaling pathway | $3.2 * 10^{-49}$ |
| | GO:0007166 | cell surface receptor linked signal transduction | $7.5 * 10^{-15}$ |
| | GO:0032502 | developmental process | $4.1 * 10^{-12}$ |
| | GO:0031293 | membrane protein intracellular domain proteolysis | $4.8 * 10^{-12}$ |
| | GO:0007389 | pattern specification process | $5.9 * 10^{-12}$ |
| | … | | |
| TGF | GO:0007178 | transmembrane receptor protein serine/threonine kinase signaling pathway | $3.4 * 10^{-54}$ |
| | GO:0007179 | transforming growth factor beta receptor signaling pathway | $4.1 * 10^{-41}$ |
| | GO:0007167 | enzyme linked receptor protein signaling pathway | $3.2 * 10^{-39}$ |
| | GO:0043687 | post-translational protein modification | $1.2 * 10^{-35}$ |
| | GO:0006464 | protein modification process | $5.5 * 10^{-33}$ |
| | … | | |
| WNT | GO:0016055 | Wnt receptor signaling pathway | $1.8 * 10^{-88}$ |
| | GO:0007166 | cell surface receptor linked signal transduction | $6.8 * 10^{-45}$ |
| | GO:0007165 | signal transduction | $3.7 * 10^{-33}$ |
| | GO:0007154 | cell communication | $1.0 * 10^{-30}$ |
| | GO:0050794 | regulation of cellular process | $3.0 * 10^{-25}$ |
| | … | | |

**Supplementary Table 12c.** The statistically most significant GO Biological Process terms for each human signaling pathway of curated in SignaLink. See the text above for details.

## Gene Ontology term similarity of human multi-pathway proteins and their pathways

As explained above, we have computed with GO TermFinder the corrected P-values for the list of significantly enriched Gene Ontology terms in each signaling pathway. To estimate the significance with which a protein belongs to a pathway a natural first step is to find GO terms that annotate both the protein and the pathway. Here we quantify the significance with which an arbitrary protein belongs to an arbitrary signaling pathway as the smallest (most significant) P-value of a GO term of the pathway that annotates the selected protein too.

In Supplementary Figure 5 we plot the results of three comparisons. First, we compute the above defined similarity for each human protein – SignaLink pathway pair, and plot the distribution of similarity scores. This is the control set shown in gray color in the figure. Next, we restrict the same comparison to protein-pathway pairs where the protein is a member of that pathway in SignaLink (red in the figure). Third, we restrict this comparison even further and use only such protein-pathway pairs, where the protein is a member of that pathway and it is a member of at least one other pathway, too, *i.e.*, it is a multi-pathway protein (see blue boxes in the figure). We found that the similarity between multi-pathway proteins and their pathways has a peak at low P-values (high significance). The same holds for the comparisons between all pathway member protein and their pathways. On the other hand, in the control case (all proteins compared to all pathways) the peak of the distribution of similarities is at high P-values (low significance).

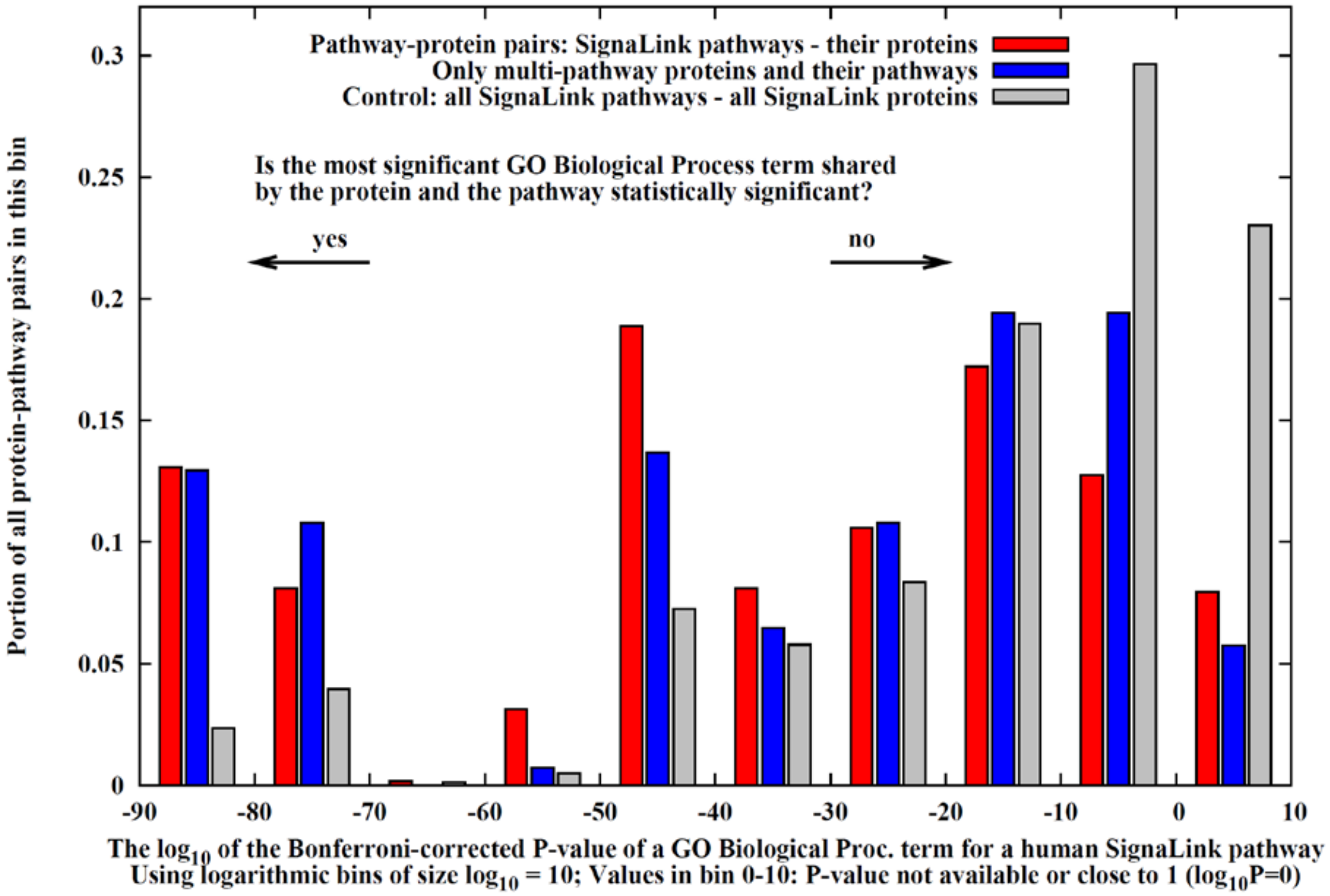

**Supplementary Figure 5.** Distribution of the similarity scores between the functions (GO Biological Process) of a protein and the functions of a SignaLink pathway. For each of the 8 human signaling pathways curated in SignaLink, GO TermFinder was used to compute the P-values of GO Biological Process terms. Given a set of proteins, their GO terms (in one of three aspects: Biological Process, Cellular component, or Molecular Function) and the DAG (Directed Acyclic Graph) formatted hierarchy of these GO terms, GO TermFinder computes the P-value of each term for the group of proteins as a whole. A summary of this analysis is shown in Supplementary Table 12. See also the text above for details. Here we found that those pathway member proteins that belong to more than one pathway are significantly more similar to their pathways than the proteins in the control case (all proteins compared to all pathways). However, they are less similar to their pathways than those proteins that belong to one pathway only, most likely because they have multiple functions. See text for further details.

## *Validation of the human interactions of SignaLink with an external resource*

We have evaluated the human interactions listed in SignaLink with the PRINCESS web service (*Protein Interaction Confidence Evaluation System with Multiple Data Sources)* (Li et al., 2008). Although PRINCESS can evaluate only human interactions, the majority of the interactions in SignaLink are human (991 from the total 1461), therefore, this analysis can provide representative information about the percent of true positive interactions in SignaLink. PRINCESS uses 7 different sources of interactions: (1) PPI databases; (2) Known interolog interactions; (3) Possibility for domain interaction; (4) GO_co-annotation; (5) Genome_context; (6) Gene_co-expression; (7) Network_topology information.

From the 991 human protein-protein interactions listed in SignaLink, the PRINCESS web service recognized 982 interactions and from these 890 interactions received a likelihood score above the cutoff threshold for reliable interactions. Thus, according to this validation test the ratio of high confidence interactions among the human signaling interactions listed in SignaLink is 90.6%. (For detailed statistics see Supplementary Table 13.) The interactions for *C. elegans* and *D. melanogaster* were compiled with the same curation rules as for human, therefore, we expect similar ratios of true positive interactions in these two species too.

| | Number of interactions |
|---|---|
| Interactions tested* (SignaLink human) | 982 |
| High-confidence interactions** | 890 |
| | |
| **Source of evidence for high-confidence interactions** | |
| PPI_Database | 201 |
| Interolog | 13 |
| Interacting_Domain | 810 |
| GO_Co-annotation | 754 |
| Genome_Context | 25 |
| Gene_Co-expression | 117 |
| Network_Topology | 451 |

**Supplementary Table 13.** Statistics of the evaluation of human interactions listed in SignaLink with the PRINCESS web service. *From the 991 human interactions listed in SignaLink, PRINCESS recognized 982 interactions. **Interactions with a likelihood ratio above 2.

## *Database updates*

The next update of the SignaLink database is scheduled for the period January, 2011 – June, 2011.

**Data sources and methods:**

- PubMed search for review papers of the curated signaling pathways.
- Semi-automatic filtering of the PubMed abstracts returned after a search for the current member proteins of SignaLink pathways.
- Publications listed (by PubMed ID) in the updates of major signaling databases.
- Search for recent high-confident high-throughput databases on protein-protein interactions of signaling proteins.

**Participants:**

- T. Korcsmaros, I. J. Farkas, T. Vellai, and P. Csermely: improving curation guidelines.
- T. Korcsmaros and D. Fazekas: curation.
- I. J. Farkas and D. Fazekas: programming.